%% file: add_force_and_or_change_projection_arxiv.tex
\documentclass[12pt]{article}
\usepackage[utf8]{inputenc}
\usepackage{graphicx}
\graphicspath{ {./images/} }
\usepackage{adjustbox}
\newcommand\myeq{\mathrel{\overset{\makebox[0pt]{\mbox{\normalfont\tiny\sffamily def}}}{=}}}
\usepackage{amsmath,amssymb}
\newtheorem{remark}{Remark}
\usepackage{url}
\usepackage{longtable}
\usepackage{caption}
\usepackage{subcaption}

\usepackage{thesis}
\usepackage{hyperref}
\usepackage[left=2.5cm,top=2.5cm,right=2.5cm,bottom=2.5cm]{geometry}

\usepackage{tocloft}

\title{Add force and/or change underlying projection method to improve accuracy of Explicit Robin-Neumann and fully decoupled schemes for the coupling of incompressible fluid with thin-walled structure}
\author{Yiyi HUANG (yhuangcm@connect.ust.hk or yiyi\_huang\_me@outlook.com)}
\providecommand{\keywords}[1]{\textbf{\textit{Keywords: }} #1}

\begin{document}

\maketitle

%
%
%

\newpage
\addcontentsline{toc}{chapter}{Table of Contents}
\tableofcontents

\newpage
\addcontentsline{toc}{chapter}{List of Figures}
\listoffigures

\newpage
\addcontentsline{toc}{chapter}{List of Tables}
\listoftables

\newpage
\addcontentsline{toc}{chapter}{Abstract}
\begin{abstract}

This work aims at providing some novel and practical ideas to improve accuracy of some partitioned algorithms, precisely Fernandez's Explicit Robin-Neumann and fully decoupled schemes, for the coupling of incompressible fluid with thin-walled structure. Inspired by viscosity of fluid and justified by boundary layer theory, the force between fluid and structure corresponding to viscosity is increased. Numerical experiments demonstrate improvement of accuracy under such modification. To improve accuracy of fully decoupled schemes further, the underlying projection method is replaced.
\end{abstract}

\newpage
\pagenumbering{arabic}
\pagestyle{plain}

\section{Background} \label{section_Background}

The coupling of incompressible fluid with thin-walled structure typically arises in bio-mechanics of blood flow in human arteries, for which the blood flow is viscous and incompressible, while the structure assumed to be deformable. The blood flow is supposed to be free of full-scale turbulence, since its Reynolds numbers in arteries are usually below $ 2000 $ \cite{cardiovascular_math}. 



Among various algorithms on this topic, Fernandez's \textbf{Explicit Robin-Neumann} \cite{rn} and \textbf{fully decoupled schemes} \cite{Fernandez_2013} \cite{Fernandez_2015}  are exceptional, due to these features:
\begin{enumerate}
  \item High efficiency. The fluid and structure are solved only once per time-step, in a genuinely partitioned style. The \textbf{fully decoupled scheme} is even more efficient than \textbf{Explicit Robin-Neumann scheme}, because the velocity and pressure of fluid are decoupled.
  \item Unconditional stability free of added-mass effect proved by theoretical analysis or numerical experiments (availability of theoretical analysis depends on extrapolation orders).
  \item First-order accuracy (under extrapolations of first- or second- order).
  \item Applicability to the topic with a vast variety of structure models.
\end{enumerate}
The other algorithms do not possess all these properties, to the best knowledge of the author of this work. Some are less efficient  (\cite{stable_loosely_Guidoboni} \cite{kinematic_splitting_Lukacova} \cite{fsi_Bukac}) or accurate (\cite{stabilization_explicit_coupling_Fernandez} \cite{coupling_schemes_Fernandez} \cite{incremental_schemes_Fernandez}), some are unstable independently of space or time parameters (e.g. the standard Dirichlet–Neumann explicit coupling scheme) or stable only under restrictive constraints on these parameters (\cite{stabilization_explicit_coupling_Fernandez} \cite{coupling_schemes_Fernandez}), and others are only applicable to the interaction of fluid with specific structure due to the way of convergence analysis (\cite{fsi_Robin_Badia} ).

\begin{remark}
  Instability of some algorithms on this topic is believed to be induced by added-mass effect (see e.g. \cite{cardiovascular_math}). At the beginning, this work was initiated by the study of added-mass effect, but does not focus on it later, so here only presents some notes. In fluid mechanics, when a body moves in a fluid, the inertia of the fluid opposes the motion, because the body and the fluid can not occupy the physical space at the same time. It is equivalent to having a \textit{virtual} mass added to the mass of the body \cite{added_mass_Abrate} \cite{added_mass_computation}. Such effect is named added mass effect. The concept was first investigated by Dubua (1776), who conducted experiments on spherical pendulum of low swings \cite{hydrodynamics_Birkhoff}. Exact mathematical expressions of added mass for a sphere were obtained by Green (1833) and Stokes (1843) respectively \cite{hydrodynamics_Lamb}. Later the concept was generalized to a desired body moving in different flow regimes \cite{added_mass_Korotkin}. Nobile \textit{et al.} \cite{added_mass_effect} analysed mathematically the added-mass effect for a linear, incompressible and inviscid fluid coupled with a linear, elastic and cylindrical tube and applied the analysis to some explicit and implicit schemes to prove their unconditional instabilities or find out the conditions of stabilities. 



\end{remark}

\section{Motivation} \label{section_Motivation}
Compared with fully implicit algorithms, partitioned algorithms are less accurate. This work aims at providing some novel and practical ideas to improve accuracy of some partitioned algorithms, precisely Fernandez's \textbf{Explicit Robin-Neumann} and \textbf{fully decoupled schemes}, for the coupling of incompressible fluid with thin-walled structure.


From point of view of physical intuition, due to viscosity of fluid, when the structure moves, some amount of fluid is attached to it near the interface, resulting in some \textit{actual} mass added to the structure. By Newton's Second Law, the mass together with acceleration produces a force on the structure; on the other hand, viscosity causes skin friction between fluid and structure. By Newton's Third Law, there is an equal and opposite force on the fluid. The application of Newton's Third Law ensures balance of force. 

The intuition can be justified by the theory of boundary layer, which was first introduced by Ludwig Prandtl in 1904 concerning the motion of a fluid with small viscosity near the wall of a solid boundary \cite{prandtl1904motion}. The theory states, adjacent to the wall, there exists a thin transition layer, within which the viscosity of fluid takes significant effect on the motion of fluid (in fact, inside the layer, the viscous force is so large that it is of the same order with inertia force). The layer is called \textit{boundary layer}. Over decades, the theory grew and was validated by an amount of experimental observations in various scientific and engineering fields \cite{history_boundary_layer_Tani}. It is also worth noting that the pressure remains \textit{almost constant} (in the senses that the pressure gradient is negligible) through the boundary layer in a direction normal to the interface. 

Thus, it is reasonable to expect, for a partitioned algorithm, increasing the force between the fluid and structure due to viscosity of fluid can make it more realistic, namely more close to the actual behaviour of the coupling system, which might lead to better accuracy. The amount of extra force resulting from viscosity is difficult (if not impossible) to compute; however, it can be approached. Up to certain values, increasing the force gradually should improve accuracy gradually.


The idea is applied to Fernandez's two schemes mentioned above. In what follows, the two schemes are presented. Afterwards, coefficients of the terms corresponding to force from viscosity are increased, generating \textbf{Algorithm 1} and \textbf{Algorithm 2}. Fernandez's \textbf{fully decoupled scheme} is based on Chorin-Temam projection method. To improve accuracy further, the underlying projection method is replaced with Van Kan's, leading to \textbf{Algorithm 3}.

Numerical results are reported later. The results indicate improvement of accuracy as the force from viscosity increases. Appropriate values for such increment are recommended for practical applications. 


\section{The simplified model problem}
\begin{figure} [h]
\centering
\includegraphics[width = 0.9\textwidth]{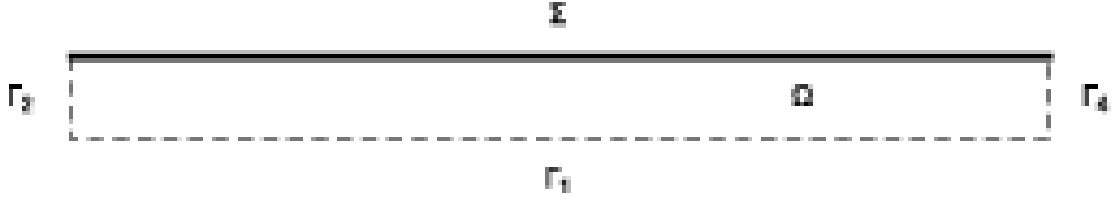}
\caption{Geometrial configuration}
\label{domain}
\end{figure}

For sake of clarity and simplicity, the simplified test-case used in \cite{tutorial1} is adopted. This work is expected to be also applicable to the a bit more general model described in \cite{rn} \cite{Fernandez_2013} \cite{Fernandez_2015}. The fluid dominated by Stokes equations is defined on $ \Omega = [0, L] \times [0, R] $, where $ L = 6, R = 0.5 $ (all the quantities are under CGS system) , with $ \partial \Omega =  \Gamma_1 \cup \Gamma_2 \cup \Sigma \cup \Gamma_4 $ (see Figure \ref{domain}  )   . The domain is extracted as upper half of the rectangle $[0, L] \times [-R, R] $ which simulates a tube in two-dimensional space with horizontal centerline $ \Gamma_1 $ and top boundary $ \Sigma $. As a result, $ \Gamma_1 $ is imposed symmetric boundary condition. The structure is assumed to be a generalized string defined on $ \Sigma $ with the two end points ( $ x = 0, L $ ) fixed. When the fluid flows from the left to the right, structure deforms vertically. Equations read as follows.

Find the fluid velocity $ \textbf{u}: \Omega \times \mathbb{R}^+ \rightarrow \mathbb{R}^2 $, the fluid pressure $ p: \Omega \times \mathbb{R}^+ \rightarrow \mathbb{R} $, the structure vertical displacement $ \eta : \Sigma \times \mathbb{R}^+  \rightarrow \mathbb{R} $ and the structure vertical velocity $ \dot{\eta} : \Sigma \times \mathbb{R}^+ \rightarrow \mathbb{R} $ such that

\begin{eqnarray}  \label{eq:fluid_model}
\left\{
\begin{aligned}
\rho^f \partial_t \boldsymbol{u} - \textbf{div } \boldsymbol{\sigma}  ( \boldsymbol{u} ,p ) & = \textbf{0}   & \text{in \quad} & \Omega,   \\
\textbf{div}  \hspace{1pt} \boldsymbol{u} &= 0   & \text{in \quad}       & \Omega ,  \\
\boldsymbol{u} \cdot \boldsymbol{n} = 0 , \boldsymbol{\sigma} ( \boldsymbol{u} , p ) \boldsymbol{n} \cdot \boldsymbol{t} & = 0   & \text{on \quad}  & \Gamma_1,   \\
 \boldsymbol{\sigma} ( \boldsymbol{u} , p ) \boldsymbol{n} & = - \textit{P} (t) \boldsymbol{n}   & \text{on \quad}  & \Gamma_2,   \\
\boldsymbol{\sigma} ( \boldsymbol{u} , p ) \boldsymbol{n}  &= \boldsymbol{0}  & \text{on \quad}  & \Gamma_4,     
\end{aligned}
\right.
\end{eqnarray}

\begin{eqnarray}  \label{eq:solid_model}
\left\{
\begin{aligned}
\boldsymbol{u} \cdot \boldsymbol{n} = \dot{\eta}, \boldsymbol{u} \cdot \boldsymbol{t} = & 0 \quad & \text{on \quad} & \Sigma,    \\
\rho^s \epsilon \partial_t \dot{\eta} - c_1 \partial_{xx} \eta + c_0 \eta = - \boldsymbol{\sigma} ( \boldsymbol{u}, p ) \boldsymbol{n} \cdot  & \boldsymbol{n} \quad & \text{on \quad} & \Sigma,  \\
\dot{\eta} = \partial_t & \eta  & \text{on \quad} & \Sigma,   \\
\eta =  & 0 \quad & \quad \text{on \quad} & \partial \Sigma, 
\end{aligned}
\right.
\end{eqnarray}

\noindent
with initial conditions 
$$ \boldsymbol{u} ( 0 ) = \boldsymbol{0}, \eta (0) = 0, \dot{\eta} (0) = 0,  $$

\noindent
where normal vector is denoted by $ \boldsymbol{n} $, tangent vector is $ \boldsymbol{t} $, 
fluid Cauthy stress tensor $ \boldsymbol{ \sigma } ( \boldsymbol{u} , p )  \myeq -p \boldsymbol{I} + 2 \mu \boldsymbol{ \varepsilon(u) } ,\quad  \boldsymbol{ \varepsilon(u) }  \myeq  \frac{1}{2} \boldsymbol{ ( \nabla u + \nabla u^T ) } $,  
fluid dynamic viscosity $ \mu = 0.035 $, fluid density $ \rho^f = 1.0$, 
pressure $ P (t) = P_{max} ( 1 - cos(2t \pi / T^\star )) /2  $, $ P_{max} = 2*10^4   \textrm{ when }  0 \leq t \leq T^\star \text{ and } P_{max} = 0  \textrm{ when }  t > T^\star $,  
$ T^\star = 5*10^{-3} , $
structure density $ \rho^s = 1.1, c_1 \myeq \frac{E \epsilon}{2 (1+ \nu) } , c_0 \myeq   \frac{E \epsilon}{R^2 (1- \nu^2) } ,  \epsilon = 0.1 $, Young's modulus $ E = 0.75*10^6 $, Poisson's ratio $ \nu = 0.5 $.

\section{Notations}
For all the algorithms mentioned in this work, $ \tau $ denotes time step, while $ h $ stands for space discretization parameter. The integer $ n $ is employed to count time steps. 

Given arbitrary variable $x $, the notation 
\begin{eqnarray} 
x^{n, \star} \myeq 
\left\{
\begin{aligned}
& 0 & \text{if \quad} & r  = 0,  \\
& x^{n-1} & \text{if \quad} & r  = 1,  \\
& 2x^{n-1} - x^{n-2} & \text{if \quad} & r = 2
\end{aligned}
\right.
\end{eqnarray}
 is used for interface extrapolations of order $ r $.

\section{Fernandez's \textbf{Explicit Robin-Neumann} and \textbf{fully decoupled schemes}}
The time semi-discrete form of \textbf{Explicit Robin-Neumann scheme} (Fernandez \cite{rn}) is cited here. 

\par\noindent\rule{\textwidth}{1pt}
\textbf{(Fernandez) Explicit Robin-Neumann scheme (time semi-discrete)} 
\par\noindent\rule{\textwidth}{.5pt}
\noindent
For $ n \geq r + 1 $,  find $ \textbf{u}^n: \Omega  \rightarrow \mathbb{R}^2 $, $ p^n: \Omega  \rightarrow \mathbb{R} $, $ \eta^n : \Sigma   \rightarrow \mathbb{R} $ and $ \dot{\eta}^n : \Sigma  \rightarrow \mathbb{R} $ such that

1. Fluid step (interface Robin condition)
\begin{eqnarray}  \label{eq:fluid_ern}
\left\{
\begin{aligned}
  \rho^f \frac{\textbf{\textbf{u}}^{n}-\textbf{u}^{n-1} }{ \tau} - \textbf{div } \boldsymbol{\sigma}  ( \boldsymbol{u}^n ,p^n ) & = \textbf{0}   & \text{in \quad} & \Omega,   \\
\textbf{div}  \hspace{1pt} \boldsymbol{u}^n &= 0   & \text{in \quad}       & \Omega ,  \\
\boldsymbol{u}^n \cdot \boldsymbol{n} = 0 , \boldsymbol{\sigma} ( \boldsymbol{u}^n , p^n ) \boldsymbol{n} \cdot \boldsymbol{t} & = 0   & \text{on \quad}  & \Gamma_1,   \\
 \boldsymbol{\sigma} ( \boldsymbol{u}^n , p^n ) \boldsymbol{n} & = - \textit{P} (t) \boldsymbol{n}   & \text{on \quad}  & \Gamma_2,   \\
\boldsymbol{\sigma} ( \boldsymbol{u}^n , p^n ) \boldsymbol{n}  &= \boldsymbol{0}  & \text{on \quad}  & \Gamma_4,     \\
\boldsymbol{\sigma} ( \boldsymbol{u}^n , p^n ) \boldsymbol{n}  \cdot \boldsymbol{n} + \frac{\rho^s \epsilon}{\tau} \boldsymbol{u}^n \cdot \boldsymbol{n} = \frac{\rho^s \epsilon}{\tau} ( \dot{\eta}^{n-1} + \tau \partial_{\tau} \dot{\eta}^{n, \star} ) &&&    \\
  + (-p^{n, \star} \boldsymbol{I} + 2 \mu \boldsymbol{ \varepsilon( \boldsymbol{u}^{n, \star} ) }) & \boldsymbol{n} \cdot   \boldsymbol{n}  & \text{on \quad}  & \Sigma,  \\
\boldsymbol{u}^n \cdot \boldsymbol{t} & = 0 & \text{on \quad} & \Sigma.
\end{aligned}
\right.
\end{eqnarray}

2. Solid step (Neumann condition)
\begin{eqnarray}  \label{eq:solid_ern}
\left\{
\begin{aligned}
  \rho^s \epsilon \frac{\dot{\eta}^n-\dot{\eta}^{n-1}}{\tau} - c_1 \partial_{xx} \eta^n + c_0 \eta^n = & -(  -p^{n} \boldsymbol{I} + 2 \mu \boldsymbol{ \varepsilon(u^n) } ) \boldsymbol{n} \cdot  \boldsymbol{n} & \text{on \quad} & \Sigma,  \\
  \dot{\eta}^n = & \partial_{\tau}  \eta^n  & \text{on \quad} & \Sigma,   \\
\eta^n =  & 0 \quad & \quad \text{on \quad} & \partial \Sigma, 
\end{aligned}
\right.
\end{eqnarray}
\par\noindent\rule{\textwidth}{.5pt}

 The \textbf{fully decoupled scheme} is proposed in Fernandez \cite{Fernandez_2013}, \cite{Fernandez_2015}. There are non-incremental and incremental forms, of which both deliver close numerical results on accuracy.  Here only presents the non-incremental form.

\par\noindent\rule{\textwidth}{1pt}
\textbf{(Fernandez) fully decoupled  scheme (time semi-discrete)} 
\par\noindent\rule{\textwidth}{.5pt}
For $ n \geq r + 1 $,

(1) Fluid viscous sub-step: find $ \textbf{\~{u}}^{n}: \Omega  \rightarrow \mathbb{R}^2 $ such that
\begin{eqnarray}  \label{eq:fluid_viscous_fully_decoupled}
\left\{
\begin{aligned}
  \rho^f \frac{\textbf{\~{u}}^{n}-\textbf{u}^{n-1} }{ \tau} - 2 \mu \textbf{div } \boldsymbol{\varepsilon} (\textbf{\~{u}}^{n}) = 0 \quad & \text{in} & \Omega,   \\
  \textbf{\~{u}}^n \cdot \boldsymbol{n} = 0 , 2 \mu \boldsymbol{\varepsilon} (\textbf{\~{u}}^{n}) \boldsymbol{n} \cdot \boldsymbol{\tau} = 0 \quad  & \text{on}  & \Gamma_1,   \\
  2 \mu \boldsymbol{\varepsilon} (\textbf{\~{u}}^{n}) \boldsymbol{n} \cdot \boldsymbol{\tau} = 0 \quad  & \text{on}  & \Gamma_2,   \\
  2 \mu \boldsymbol{\varepsilon} (\textbf{\~{u}}^{n}) \boldsymbol{n} \cdot \boldsymbol{\tau} = 0 \quad  & \text{on}  & \Gamma_4,   \\
  \tilde{u}_1^n = 0, 2 \mu \boldsymbol{\varepsilon} (\textbf{\~{u}}^{n}) \boldsymbol{n} \cdot \boldsymbol{n} + \frac{\rho^s \epsilon}{\tau} \textbf{\~{u}}^n \cdot \boldsymbol{n} =   \frac{\rho^s \epsilon}{\tau} \dot{\eta}^{n-1} & \text{on} & \Sigma, \\
\end{aligned}
\right.
\end{eqnarray}

(2) Fluid projection sub-step: find $ \phi^n: \Omega  \rightarrow \mathbb{R}^2 $ such that
\begin{eqnarray}  \label{eq:fluid_projection_fully_decoupled}
\left\{
\begin{aligned}
  -\frac{\tau}{\rho^f} \Delta \phi^n = -\nabla \cdot \textbf{\~{u}}^n \quad & \text{in} & \Omega, \\
  \frac{\partial \phi^n}{\partial n} = 0 \quad & \text{on} & \Gamma_1, \\
  \phi^n = P(t_n) \quad & \text{on} & \Gamma_2, \\
  \phi^n = 0 \quad & \text{on} & \Gamma_4,  \\
  \frac{\tau}{\rho^f} \nabla \phi^n \cdot \boldsymbol{n} + \frac{\tau}{\rho^s \epsilon} \phi^n = \frac{\tau}{\rho^s \epsilon} \phi^{n,\star} + \textbf{\~{u}}^{n,\star} \cdot \boldsymbol{n} - \dot{\eta}^{n,\star}  \quad & \text{on} & \Sigma,  \\
\end{aligned}
\right.
\end{eqnarray}
Thereafter set $ p^n = \phi^n, \boldsymbol{u}^n = \textbf{\~{u}}^n - \frac{\tau}{\rho^f} \nabla \phi ^n \quad  \text{in} \quad \Omega . $ 

(3) Solid sub-step: find $ \eta^n: \Sigma \rightarrow \mathbb{R}^2 $ such that
\begin{eqnarray}  \label{eq:solid_fully_decoupled}
\left\{
\begin{aligned}
  \rho^s \epsilon \frac{\dot{\eta}^n-\dot{\eta}^{n-1}}{\tau} - c_1 \partial_{xx} \eta^n + c_0 \eta^n = -2 \mu \boldsymbol{\varepsilon} (\textbf{\~{u}}^n) \textbf{n} \cdot \textbf{n} + p^n \quad & \text{on} & \Sigma,  \\
  \dot{\eta}^n = \frac{\eta^n-\eta^{n-1}}{\tau}  \quad & \text{on} & \Sigma,   \\
  \eta^n =   0 \quad & \text{on} & \partial \Sigma, 
\end{aligned}
\right.
\end{eqnarray}
\par\noindent\rule{\textwidth}{.5pt}

\begin{remark}
  Substituting  $ \boldsymbol{u}^n = \textbf{\~{u}}^n - \frac{\tau}{\rho^f} \nabla \phi ^n \text{ into } \eqref{eq:fluid_viscous_fully_decoupled}_{1}  $ leads to a more compact style with $ \boldsymbol{u}^n $ eliminated (see \cite{Fernandez_2013}).
\end{remark}

\section{Two schemes with added force}
Replacing the coefficient $ 2 $ in the term $ 2 \mu \boldsymbol{ \varepsilon( \boldsymbol{u}^{n, \star} ) } $ at the right hand side of $ \eqref{eq:fluid_ern}_{6} $ and the term $ 2 \mu \boldsymbol{ \varepsilon( \boldsymbol{u}^{n} ) } $ at the right hand side of $ \eqref{eq:solid_ern}_{1} $ with a real number named $ \beta $ larger than 2 generates \textbf{Algorithm 1} as follows. Forces are balanced on the interface under such modifications. 

Analogously, substituting the coefficient $ 2 $ in the term $ -2 \mu \boldsymbol{\varepsilon} (\textbf{\~{u}}^n) \textbf{n} \cdot \textbf{n} $ at the right hand side of $ \eqref{eq:solid_fully_decoupled}_{1} $ with a real number named $ \theta $ larger than $ 2 $ leads to \textbf{Algorithm 2}. 

\begin{remark}
  There is no such a term in $ \eqref{eq:fluid_viscous_fully_decoupled}_{5} $ like $  (-p^{n, \star} \boldsymbol{I} + 2 \mu \boldsymbol{ \varepsilon( \boldsymbol{u}^{n, \star} ) })  \boldsymbol{n} \cdot   \boldsymbol{n}   $ of $ \eqref{eq:fluid_ern}_{6} $, so there is no such a term in $ \eqref{eq:fluid_viscous_algorithm_2}_{5} $ like $  (-p^{n, \star} \boldsymbol{I} + \beta \mu \boldsymbol{ \varepsilon( \boldsymbol{u}^{n, \star} ) })  \boldsymbol{n} \cdot   \boldsymbol{n}   $ of $ \eqref{eq:fluid_algorithm_1}_{6} $. To \textbf{Fernandez's fully decoupled scheme} or \textbf{Algorithm 2}, balance of forces on the interface may be understood in the way that some terms disappear during the underlying procedure of operator splitting (note that the two schemes are based on projection methods, which are essentially some kinds of operator-splitting schemes), which makes it less evident as that of \textbf{Fernandez's Explicit Robin-Neumann scheme} or \textbf{Algorithm 1}. The same argument should apply to the coming \textbf{Algorithm 3} regarding the balance of force. 
\end{remark}

\begin{remark}
  The coefficient of pressure on the interface is not augmented, because of the reason stated in \textbf{Section \ref{section_Motivation}}.
\end{remark}

\par\noindent\rule{\textwidth}{1pt}
\textbf{Algorithm 1} 
\par\noindent\rule{\textwidth}{.5pt}
\noindent
For real number $ \beta > 2, n \geq r + 1 $,  find $ \textbf{u}^n: \Omega  \rightarrow \mathbb{R}^2 $, $ p^n: \Omega  \rightarrow \mathbb{R} $, $ \eta^n : \Sigma   \rightarrow \mathbb{R} $ and $ \dot{\eta}^n : \Sigma  \rightarrow \mathbb{R} $ such that

1. Fluid step (interface Robin condition)
\begin{eqnarray}  \label{eq:fluid_algorithm_1}
\left\{
\begin{aligned}
  \rho^f \frac{\textbf{\textbf{u}}^{n}-\textbf{u}^{n-1} }{ \tau} - \textbf{div } \boldsymbol{\sigma}  ( \boldsymbol{u}^n ,p^n ) & = \textbf{0}   & \text{in \quad} & \Omega,   \\
\textbf{div}  \hspace{1pt} \boldsymbol{u}^n &= 0   & \text{in \quad}       & \Omega ,  \\
\boldsymbol{u}^n \cdot \boldsymbol{n} = 0 , \boldsymbol{\sigma} ( \boldsymbol{u}^n , p^n ) \boldsymbol{n} \cdot \boldsymbol{t} & = 0   & \text{on \quad}  & \Gamma_1,   \\
 \boldsymbol{\sigma} ( \boldsymbol{u}^n , p^n ) \boldsymbol{n} & = - \textit{P} (t) \boldsymbol{n}   & \text{on \quad}  & \Gamma_2,   \\
\boldsymbol{\sigma} ( \boldsymbol{u}^n , p^n ) \boldsymbol{n}  &= \boldsymbol{0}  & \text{on \quad}  & \Gamma_4,     \\
\boldsymbol{\sigma} ( \boldsymbol{u}^n , p^n ) \boldsymbol{n}  \cdot \boldsymbol{n} + \frac{\rho^s \epsilon}{\tau} \boldsymbol{u}^n \cdot \boldsymbol{n} = \frac{\rho^s \epsilon}{\tau} ( \dot{\eta}^{n-1} + \tau \partial_{\tau} \dot{\eta}^{n, \star} ) &&&    \\
  + (-p^{n, \star} \boldsymbol{I} + \beta \mu \boldsymbol{ \varepsilon( \boldsymbol{u}^{n, \star} ) }) & \boldsymbol{n} \cdot   \boldsymbol{n}  & \text{on \quad}  & \Sigma,  \\
\boldsymbol{u}^n \cdot \boldsymbol{t} & = 0 & \text{on \quad} & \Sigma.
\end{aligned}
\right.
\end{eqnarray}

2. Solid step (Neumann condition)
\begin{eqnarray}  \label{eq:solid_algorithm_1}
\left\{
\begin{aligned}
  \rho^s \epsilon \frac{\dot{\eta}^n-\dot{\eta}^{n-1}}{\tau} - c_1 \partial_{xx} \eta^n + c_0 \eta^n = & -(  -p^{n} \boldsymbol{I} + \beta \mu \boldsymbol{ \varepsilon(u^{n}) } ) \boldsymbol{n} \cdot  \boldsymbol{n} & \text{on \quad} & \Sigma,  \\
  \dot{\eta}^n = & \partial_{\tau}  \eta^n  & \text{on \quad} & \Sigma,   \\
\eta^n =  & 0 \quad & \quad \text{on \quad} & \partial \Sigma, 
\end{aligned}
\right.
\end{eqnarray}
\par\noindent\rule{\textwidth}{.5pt}

\par\noindent\rule{\textwidth}{1pt}
\textbf{Algorithm 2} 
\par\noindent\rule{\textwidth}{.5pt}
For real number $ \theta > 2 $, $ n \geq r + 1 $,  

(1) Fluid viscous sub-step: find $ \textbf{\~{u}}^{n}: \Omega  \rightarrow \mathbb{R}^2 $ such that
\begin{eqnarray}  \label{eq:fluid_viscous_algorithm_2}
\left\{
\begin{aligned}
  \rho^f \frac{\textbf{\~{u}}^{n}-\textbf{u}^{n-1} }{ \tau} - 2 \mu \textbf{div } \boldsymbol{\varepsilon} (\textbf{\~{u}}^{n}) = 0 \quad & \text{in} & \Omega,   \\
  \textbf{\~{u}}^n \cdot \boldsymbol{n} = 0 , 2 \mu \boldsymbol{\varepsilon} (\textbf{\~{u}}^{n}) \boldsymbol{n} \cdot \boldsymbol{\tau} = 0 \quad  & \text{on}  & \Gamma_1,   \\
  2 \mu \boldsymbol{\varepsilon} (\textbf{\~{u}}^{n}) \boldsymbol{n} \cdot \boldsymbol{\tau} = 0 \quad  & \text{on}  & \Gamma_2,   \\
  2 \mu \boldsymbol{\varepsilon} (\textbf{\~{u}}^{n}) \boldsymbol{n} \cdot \boldsymbol{\tau} = 0 \quad  & \text{on}  & \Gamma_4,   \\
  \tilde{u}_1^n = 0, 2 \mu \boldsymbol{\varepsilon} (\textbf{\~{u}}^{n}) \boldsymbol{n} \cdot \boldsymbol{n} + \frac{\rho^s \epsilon}{\tau} \textbf{\~{u}}^n \cdot \boldsymbol{n} =   \frac{\rho^s \epsilon}{\tau} \dot{\eta}^{n-1} & \text{on} & \Sigma, \\
\end{aligned}
\right.
\end{eqnarray}

(2) Fluid projection sub-step: find $ \phi^n: \Omega  \rightarrow \mathbb{R}^2 $ such that
\begin{eqnarray}  \label{eq:fluid_projection_algorithm_2}
\left\{
\begin{aligned}
  -\frac{\tau}{\rho^f} \Delta \phi^n = -\nabla \cdot \textbf{\~{u}}^n \quad & \text{in} & \Omega, \\
  \frac{\partial \phi^n}{\partial n} = 0 \quad & \text{on} & \Gamma_1, \\
  \phi^n = P(t_n) \quad & \text{on} & \Gamma_2, \\
  \phi^n = 0 \quad & \text{on} & \Gamma_4,  \\
  \frac{\tau}{\rho^f} \nabla \phi^n \cdot \boldsymbol{n} + \frac{\tau}{\rho^s \epsilon} \phi^n = \frac{\tau}{\rho^s \epsilon} \phi^{n,\star} + \textbf{\~{u}}^{n,\star} \cdot \boldsymbol{n} - \dot{\eta}^{n,\star}  \quad & \text{on} & \Sigma,  \\
\end{aligned}
\right.
\end{eqnarray}
Thereafter set $ p^n = \phi^n, \boldsymbol{u}^n = \textbf{\~{u}}^n - \frac{\tau}{\rho^f} \nabla \phi ^n \quad  \text{in} \quad \Omega . $ 

(3) Solid sub-step: find $ \eta^n: \Sigma \rightarrow \mathbb{R}^2 $ such that
\begin{eqnarray}  \label{eq:solid_algorithm_2}
\left\{
\begin{aligned}
  \rho^s \epsilon \frac{\dot{\eta}^n-\dot{\eta}^{n-1}}{\tau} - c_1 \partial_{xx} \eta^n + c_0 \eta^n = -\theta \mu \boldsymbol{\varepsilon} (\textbf{\~{u}}^n) \textbf{n} \cdot \textbf{n} + p^n \quad & \text{on} & \Sigma,  \\
  \dot{\eta}^n = \frac{\eta^n-\eta^{n-1}}{\tau}  \quad & \text{on} & \Sigma,   \\
  \eta^n =   0 \quad & \text{on} & \partial \Sigma, 
\end{aligned}
\right.
\end{eqnarray}
\par\noindent\rule{\textwidth}{.5pt}

\section{A fully decoupled scheme based on Van Kan's projection method and with added force}
Fernandez's \textbf{fully decoupled scheme} is based on Chorin-Temam projection method, whose accuracy is of first order in time (see e.g. \cite{kuzmin} \cite{projection_overview} ). It is expected, if the underlying projection method is replaced with Van Kan's projection method, which is of second order in time (see e.g. \cite{kuzmin}), such schemes could be more accurate. This idea produces \textbf{Algorithms 3}. 

\par\noindent\rule{\textwidth}{1pt}
\textbf{Algorithm 3} 
\par\noindent\rule{\textwidth}{.5pt}
For  real number $ \xi \geq 2 $, $ n \geq r + 1 $, 

(1) Fluid viscous sub-step: find $ \textbf{\~{u}}^{n}: \Omega  \rightarrow \mathbb{R}^2 $ such that
\begin{eqnarray}  
\left\{
\begin{aligned}
  \rho^f \frac{\textbf{\~{u}}^{n}-\textbf{u}^{n-1} }{ \tau} = -\nabla{p}^{n-1} +\frac{1}{2}( 2 \mu \textbf{div } \boldsymbol{\varepsilon} (\textbf{\~{u}}^{n})  + 2 \mu \textbf{div } \boldsymbol{\varepsilon} (\textbf{u}^{n-1}) ) \quad & \text{in} & \Omega,   \\
  \textbf{\~{u}}^n \cdot \boldsymbol{n} = 0 , 2 \mu \boldsymbol{\varepsilon} (\textbf{\~{u}}^{n}) \boldsymbol{n} \cdot \boldsymbol{\tau} = 0 \quad  & \text{on}  & \Gamma_1,   \\
  2 \mu \boldsymbol{\varepsilon} (\textbf{\~{u}}^{n}) \boldsymbol{n} \cdot \boldsymbol{\tau} = 0 \quad  & \text{on}  & \Gamma_2,   \\
  2 \mu \boldsymbol{\varepsilon} (\textbf{\~{u}}^{n}) \boldsymbol{n} \cdot \boldsymbol{\tau} = 0 \quad  & \text{on}  & \Gamma_4,   \\
  \tilde{u}_1^n = 0, 2 \mu \boldsymbol{\varepsilon} (\textbf{\~{u}}^{n}) \boldsymbol{n} \cdot \boldsymbol{n} + \frac{\rho^s \epsilon}{\tau} \textbf{\~{u}}^n \cdot \boldsymbol{n} =   \frac{\rho^s \epsilon}{\tau} \dot{\eta}^{n-1} & \text{on} & \Sigma, \\
\end{aligned}
\right.
\end{eqnarray}

(2) Fluid projection sub-step: find $ \phi^n: \Omega  \rightarrow \mathbb{R}^2 $ such that
\begin{eqnarray}  \label{eq:fluid_projection_algorithm_3}
\left\{
\begin{aligned}
  -\frac{\tau}{\rho^f} \Delta \phi^n = -\nabla \cdot \textbf{\~{u}}^n \quad & \text{in} & \Omega, \\
  \frac{\partial \phi^n}{\partial n} = 0 \quad & \text{on} & \Gamma_1, \\
  \phi^n = \frac{P(t_n) - P(t_{n-1})}{2}  \quad & \text{on} & \Gamma_2, \\
  \phi^n = 0 \quad & \text{on} & \Gamma_4,  \\
  \frac{\tau}{\rho^f} \nabla \phi^n \cdot \boldsymbol{n} + \frac{\tau}{\rho^s \epsilon} \phi^n = & & \\
  \frac{\tau}{\rho^s \epsilon} \frac{p^{n,\star}-p^{n-1,\star}}{2} + \frac{ \textbf{\~{u}}^{n,\star}-\textbf{\~{u}}^{n-1,\star} }{2} \cdot \boldsymbol{n} - \frac{ \dot{\eta}^{n,\star} - \dot{\eta}^{n-1,\star} }{2}  \quad & \text{on} & \Sigma,  \\
\end{aligned}
\right.
\end{eqnarray}
Thereafter set $ p^n = p^{n-1} + 2 \phi^n, \boldsymbol{u}^n = \textbf{\~{u}}^n - \frac{\tau}{\rho^f} \nabla \phi ^n \quad  \text{in } \Omega . $ 

(3) Solid sub-step: find $ \eta^n: \Sigma \rightarrow \mathbb{R}^2 $ such that
\begin{eqnarray}  \label{eq:solid_2}
\left\{
\begin{aligned}
  \rho^s \epsilon \frac{\dot{\eta}^n-\dot{\eta}^{n-1}}{\tau} - c_1 \partial_{xx} \eta^n + c_0 \eta^n = -\xi \mu \boldsymbol{\varepsilon} (\textbf{\~{u}}^n) \textbf{n} \cdot \textbf{n} + p^n \quad & \text{on} & \Sigma,  \\
  \dot{\eta}^n = \frac{\eta^n-\eta^{n-1}}{\tau}  \quad & \text{on} & \Sigma,   \\
  \eta^n =   0 \quad & \text{on} & \partial \Sigma, 
\end{aligned}
\right.
\end{eqnarray}
\par\noindent\rule{\textwidth}{.5pt}

\begin{remark}
  Boundary conditions for the fluid projection sub-step of \textbf{Algorithm 3} are deduced from that of \textbf{Fernandez's fully decoupled scheme} by noting that $\phi^{n}=\frac{p^{n}-p^{n-1}}{2}$ for \textbf{Algorithm 3} and that $p^{n}=\phi^{n}$ for \textbf{Fernandez's fully decoupled scheme}. For example, on $\Gamma_{2},\eqref{eq:fluid_projection_fully_decoupled}_{3} $ indicates
  \[
    \begin{aligned}
      p^{n} &=P\left(t_{n}\right) \\
      p^{n-1} &=P\left(t_{n-1}\right)
    \end{aligned}
  \]
  Taking the difference and didvided by 2 yields $ \eqref{eq:fluid_projection_algorithm_3}_{3} $
  \[
    \phi^{n}=\frac{p^{n}-p^{n-1}}{2}=\frac{P\left(t_{n}\right)-P\left(t_{n-1}\right)}{2}
  \]
  The same procedure applies to the deduction of $\eqref{eq:fluid_projection_algorithm_3}_{2,4,5}$
\end{remark}

\section{Numerical experiments}
Fernandez's two algorithms and \textbf{Algorithms 1-3} are all discretized with Galerkin finite element method in space and implemented with FreeFem++ \cite{ff++} using Lagrange $ P_{1} $ element for both the fluid and structure with symmetric pressure stabilization method \cite{stabilization}. In order to observe the order of convergence, the time and space are refined at the same \textit{rate},

$$  (\tau, h ) = \frac{ (5*10^{-4}, 0.1) }{ 2^{rate} }, \quad rate = 0, 1, 2, 3, 4, 5, ... . $$

The reference solution is generated using monolithic scheme at high time-space grid resolution $ \tau = 10^{-6}, h = 3.125 \times 10^{-3}  $.  All algorithms run from initial time $ t = 0 $ to final time $ t = 0.015 $. By comparing solutions of the above $ 5 $ schemes to reference solution, relative errors in elastic energy norm (see \cite{rn}) are computed for structure displacement at final time  corresponding to different \textit{rates} of space and time refinement. 

Computation of relative errors and preparation of data for writing are completed with Perl \cite{perl} as well as an amount of Perl modules \cite{perl_modules} and Bash \cite{bash}. Graphs are drew using gnuplot \cite{gnuplot}. All codes run on x86\_64 Linux 5.6.0 \cite{linux_kernel} with one Intel\textsuperscript{\textregistered} Xeon\textsuperscript{\textregistered} E-2186M  CPU @ 2.90GHz.

Tables \ref{table:rn_m1}, \ref{table:prj_prs-coret_v3_m2} and \ref{table:prj_prs-coret_Van-Kan_m3} report relative errors of Fernandez's two algorithms and \textbf{Algorithms 1-3} with $ \beta, \theta \text{ and } \xi $ ranging from integers $ 10 $ to $ 45 $ respectively, at refinement $ \textit{rate} = 2, 3, 4, 5 $. The refinement $ \textit{rate} = 0 \text{ and } 1 $ are of no interest and not presented, since all of Fernandez's two algorithms and \textbf{Algorithms 1-3} perform poorly in accuracy at such low \textit{rates}. Numerical results of \textbf{Algorithms 1-3} with $ \beta, \theta \text{ and } \xi $ ranging from $ 2 $ to $ 10 $ are not presented, because they do not yield obvious improvement of accuracy at these intervals. 

Both of Fernandez's two algorithms achieve both highest accuracy and optimal first-order convergence rate in time with first-order extrapolation, so Tables \ref{table:rn_m1}, \ref{table:prj_prs-coret_v3_m2} and \ref{table:prj_prs-coret_Van-Kan_m3} include their results at first-order extrapolation only. For purpose of comparison, \textbf{Algorithm 1} and \textbf{2} are also computed with first-order extrapolation. However, \textbf{Algorithm 3} reaches highest accuracy at zeroth-order extrapolation, so its results at zeroth-order extrapolation are presented. 


\clearpage
\begin{center}
\input{tables/rn_m1.tex}
\end{center}

\clearpage
\begin{center}
\input{tables/prj_prs-coret_v3_m2.tex}
\end{center}

\clearpage
\begin{center}
\input{tables/prj_prs-coret_Van-Kan_m3.tex}
\end{center}

\section{Conclusions from numerical results}  \label{Conclusions}
Conclusions can be drawn from Tables \ref{table:rn_m1}, \ref{table:prj_prs-coret_v3_m2} and \ref{table:prj_prs-coret_Van-Kan_m3} respectively as follows.

\subsection{Conclusions for \textbf{Algorithm 1} from Table \ref{table:rn_m1} }
Relative errors of \textbf{Algorithm 1} decrease in a regular manner as $\beta$ or refinement \textit{rate} increase. All the relative errors are less than that of \textbf{Fernandez Explicit Robin-Neumann scheme} except for unstable ones. All algorithms roughly achieve the same convergence order in time, namely $ \mathcal{O}(t) $. 

Stability is conditional. For a specific value of $ \beta $, the algorithm is stable at high refinement \textit{rates}; for a specific refinement \textit{rate}, it is stable at small values of $ \beta $. At $ \textit{rate} = 2 $ and $ 3 $, \textbf{Algorithm 1} is stable up to $ \beta = 24 $ and $ 43 $ respectively. At $ \textit{rate} = 4 $ and $ 5 $, it is stable for all tested values of $ \beta $.

That relative errors keep decreasing as $\beta$ increases up to $ 45 $ implies that the amount of force added to the algorithm keeps approaching the actual amount of force resulting from viscosity in the coupling system. Based on the intuition mentioned in \textbf{Section \ref{section_Motivation}}, it is guessed continuing increasing $ \beta $ up to certain value larger than $ 45 $ might decrease the relative errors further. However, since the algorithm performs worse in stability at larger $ \beta $ and the stability is already frustrating at $ \beta = 45 $, it is not worth doing so.

\subsection{Conclusions for \textbf{Algorithm 2} from Table \ref{table:prj_prs-coret_v3_m2} }
Relative errors of \textbf{Algorithm 2} decrease in a regular manner as $\theta$ or refinement \textit{rate} increase.  All the relative errors are less than that of \textbf{Fernandez fully decoupled scheme} except for unstable ones.  All algorithms roughly achieve the same convergence order in time, namely $ \mathcal{O}(t) $. 

Stability is conditional. For a specific value of $ \theta $, the algorithm is stable at high refinement \textit{rates}; for a specific refinement \textit{rate}, it is stable at small values of $ \theta $. At $ \textit{rate} = 2, 3 $ and $ 4 $, \textbf{Algorithm 2} is stable up to $ \theta = 15 $ , $ 25 $ and $ 42 $ respectively. At $ \textit{rate} = 5 $ , it is stable for all tested values of $ \theta $.

That relative errors keep decreasing as $\theta$ increases up to $ 45 $ implies that the amount of force added to the algorithm keeps approaching the actual amount of force resulting from viscosity in the coupling system. Based on the intuition mentioned in \textbf{Section \ref{section_Motivation}}, it is guessed continuing increasing $ \theta $ up to certain value larger than $ 45 $ might decrease the relative errors further. However, since the algorithm performs worse in stability at larger $ \theta $ and the stability is already frustrating at $ \theta = 45 $, it is not worth doing so.

\subsection{Conclusions for \textbf{Algorithm 3} from Table \ref{table:prj_prs-coret_Van-Kan_m3} }
Relative errors of \textbf{Algorithm 3} decrease in a regular manner as $\xi$ or refinement \textit{rate} increase up to $ \xi = 27 \text{ at } \textit{rate} = 2 $, $ \xi = 31 \text{ at } \textit{rate} = 3 $, $ \xi = 33 \text{ at } \textit{rate} = 4 $ and $ \xi = 33 \text{ at } \textit{rate} = 5 $.  For larger $ \xi $ at that refinement \textit{rate}, relative errors augment. All the relative errors are less than that of \textbf{Fernandez fully decoupled scheme} except for unstable ones.  All algorithms roughly achieve the same convergence order in time, namely $ \mathcal{O}(t) $. 

Stability is conditional. For a specific value of $ \xi $, the algorithm is stable at high refinement \textit{rates}; for a specific refinement \textit{rate}, it is stable at small values of $ \xi $. At $ \textit{rate} = 2 $, \textbf{Algorithm 3} is stable up to $ \xi = 38. $ At $ \textit{rate} = 3, 4 \text{ and } 5 $ , it is stable for all tested values of $ \xi $. Compared with \textbf{Algorithm 2}, \textbf{Algorithm 3} possesses better stability and accuracy. 

That relative errors keep decreasing as $\xi$ increases up to $ 27 $ and increasing as $ \xi $ increases from $ 33 $ complies with the intuition mentioned in \textbf{Section \ref{section_Motivation}}. It is guessed the amount of force added to the algorithm obtained by taking $ \xi $ between $ 27 $ and $ 33 $ is close to the actual amount of force resulting from viscosity in the coupling system. 

\section{A possible and non-rigorous explanation for the behaviour of stability} \label{Explanation}
Theoretical analysis is not available yet. Here states a possible and non-rigorous explanation. It remains unknown whether such an explanation is correct.

All algorithms mentioned lead to linear equations after space discretizations at each time step. Compared with Fernandez's two algorithms, \textbf{Algorithms 1-3} modify the right hand side of those linear equations generated, causing perturbations to the solutions. Since Fernandez's two algorithms are stable, it is expected solutions still exist and do not change obviously under small perturbations. However, as $ \beta, \theta \text{ or } \xi $ increases, such perturbations become more significant and affect the existence and values of solutions more seriously. As time steps go on, perturbations accumulate and at some time steps cause the solutions to the linear equations generated at that step non-existent. This perhaps explains why \textbf{Algorithms 1-3} become unstable at large $ \beta, \theta \text{ or } \xi $ for a specific refinement \textit{rate}.

For a specific value of $ \beta, \theta \text{ or } \xi $, as the refinement \textit{rate} increases, the number of nodes of mesh enlarges. Let an integer $ m $ denote the number of nodes on $ \Sigma $, $ m = L/h + 1 $. The number of nodes on $ \Omega $ is approximately $ m^2 $. Note that the added force is only imposed on the interface. Thus, only $ m $ nodes are affected. The ratio of affected and non-affected nodes is approximately $ m/(m^2-m) = 1/(m-1) $, which decreases as $ m $ increases. As refinement \textit{rate} increases, $ m $ increases and therefore the forces added to \textbf{Algorithms 1-3} disturbs the system less, which yields better stability.

\section{Selection of $\beta, \theta \text{ and } \xi $ for practical applications}
Practical applications should take into account both efficiency and accuracy. At refinement $ rate = 2 $, time step $ \tau = 0.000125 $, space discretization parameter $ h = 0.025 $, it takes no more than $ 20 $ seconds to finish computation for any of Fernandez's two algorithms and \textbf{Algorithms 1-3} regardless of values of $ \beta, \theta \text{ and } \xi $. It is quite fast. However, all of them are far from accurate. Therefore, practical applications are not expected to run at such low \textit{rate} of refinement; it suffices to consider $ \textit{rate} = 3, 4 \text{ and } 5 $.

Comparing relative errors corresponding to different values at $ \textit{rate} = 3, 4 \text{ and } 5 $, the values $ \beta = 43, \theta = 25, 31 \leq \xi  \leq 33 $ are recommended for \textbf{Algorithm 1-3} respectively. Tables \ref{table:rn_m1_selected}, \ref{table:prj_prs-coret_v3_m2_selected} and \ref{table:prj_prs-coret_Van-Kan_m3_selected} report their values and percents of decrement of relative errors compared with Fernandez's algorithms respectively. The values of decrement of relative errors equal to relative errors of Fernandez's algorithms minus that of \textbf{Algorithm 1-3}, while percents equal to values divided by relative errors of Fernandez's algorithms times $100$. Structure displacements are displayed in Figures \ref{fig:rn_m1_r1_beta43}, \ref{fig:prj_prs-coret_v3_m2_r1_beta25}, \ref{fig:prj_prs-coret_Van-Kan_m3_r0_beta31}, \ref{fig:prj_prs-coret_Van-Kan_m3_r0_beta32} and \ref{fig:prj_prs-coret_Van-Kan_m3_r0_beta33}. 

\begin{center}
\input{tables/rn_m1_selected.tex}
\end{center}

\begin{center}
\input{tables/prj_prs-coret_v3_m2_selected.tex}
\end{center}

\clearpage
\begin{center}
\input{tables/prj_prs-coret_Van-Kan_m3_selected.tex}
\end{center}

\begin{center}
  \input{./images/rn_m1_r1_beta43.tex}
\end{center}

\begin{center}
  \input{./images/prj_prs-coret_v3_m2_r1_beta25.tex}
\end{center}

\begin{center}
  \input{./images/prj_prs-coret_Van-Kan_m3_r0_beta31.tex}
\end{center}

\begin{center}
  \input{./images/prj_prs-coret_Van-Kan_m3_r0_beta32.tex}
\end{center}

\begin{center}
  \input{./images/prj_prs-coret_Van-Kan_m3_r0_beta33.tex}
\end{center}

\section{Discussions and future work}
The numerical results validate the ideas that adding force corresponding to viscosity and replacing underlying projection method can improve accuracy; particularly, Table \ref{table:prj_prs-coret_Van-Kan_m3_selected} indicates as large improvement as up to $ 49.5229\% $ for \textbf{Algorithm 3} with $ \xi = 33 $ compared with \textbf{Fernandez fully decoupled scheme} at refinement $ \textit{rate} = 5 $.  It is expected, for other fluid-structure interaction problems, if the fluid is also viscous, adding force might also help with accuracy. 

As a direction of future work, it is worth trying investigating how adding force improve accuracy theoretically. Reading some works on boundary layer theory might benefit such analysis.

This work deals with accuracy. On the other hand, it is possible to improve efficiency by parallelism. A choice is to take advantage of extrapolation ($ 1^{st} $ order might be better than $0^{th}$ ). To implement such ideas, MPI \cite{mpi} might work. 






\section{Acknowledgments}
This research did not receive any specific grant from funding agencies in the public, commercial, or not-for-profit sectors.


\clearpage
\bibliographystyle{plain}
\bibliography{ref}  

\end{document}

%% file: tables/rn_m1.tex
\begin{longtable}{lllllll }
\caption{Numerical results of Fernandez Explicit Robin-Neumann scheme (Fern ERN) and Algorithm 1 (Algo 1)}
\label{table:rn_m1}\\
\hline
\hline
rate &  Fern ERN  & Algo 1  & Algo 1  & Algo 1  & Algo 1  & Algo 1  \\
     &                 &  $\beta =$ 10 & $\beta =$ 11 & $\beta =$ 12 & $\beta =$ 13 & $\beta =$ 14 \\
\hline
2 &  0.435176 & 0.423118 & 0.421689 & 0.420281 & 0.418895 & 0.417532 \\
3 &  0.241766 & 0.233158 & 0.232109 & 0.231066 & 0.230030 & 0.229001 \\
4 &  0.128616 & 0.123319 & 0.122668 & 0.122021 & 0.121377 & 0.120735 \\
5 &  0.064847 & 0.061810 & 0.061437 & 0.061066 & 0.060696 & 0.060328 \\
\hline
&&&&&&\\
&&&&&&\\
\hline
rate  & Algo 1  & Algo 1  & Algo 1  & Algo 1  & Algo 1  & Algo 1  \\
      & $\beta =$ 15 & $\beta =$ 16 & $\beta =$ 17 & $\beta =$ 18 & $\beta =$ 19 & $\beta =$ 20 \\
\hline
2 &  0.416193 & 0.414879 & 0.413592 & 0.412334 & 0.411104 & 0.409906 \\
3 &  0.227979 & 0.226965 & 0.225959 & 0.224962 & 0.223972 & 0.222992 \\
4 &  0.120097 & 0.119462 & 0.118831 & 0.118202 & 0.117578 & 0.116957 \\
5 &  0.059961 & 0.059597 & 0.059234 & 0.058873 & 0.058514 & 0.058156 \\
\hline
&&&&&&\\
&&&&&&\\
\hline
rate  & Algo 1  & Algo 1  & Algo 1  & Algo 1  & Algo 1  & Algo 1  \\
      & $\beta =$ 21 & $\beta =$ 22 & $\beta =$ 23 & $\beta =$ 24 & $\beta =$ 25 & $\beta =$ 26 \\
\hline
2 &  0.408739 & 0.407606 & 0.406510 & 0.405836 & unstable & unstable \\
3 &  0.222021 & 0.221059 & 0.220107 & 0.219165 & 0.218234 & 0.217313 \\
4 &  0.116340 & 0.115727 & 0.115117 & 0.114512 & 0.113911 & 0.113314 \\
5 &  0.057802 & 0.057449 & 0.057097 & 0.056748 & 0.056402 & 0.056057 \\
\hline
\clearpage
&&&&&&\\
&&&&&&\\
\hline
rate  & Algo 1  & Algo 1  & Algo 1  & Algo 1  & Algo 1  & Algo 1  \\
      & $\beta =$ 27 & $\beta =$ 28 & $\beta =$ 29 & $\beta =$ 30 & $\beta =$ 31 & $\beta =$ 32 \\
\hline
2 &  unstable & unstable & unstable & unstable & unstable & unstable \\
3 &  0.216403 & 0.215505 & 0.214619 & 0.213744 & 0.212883 & 0.212034 \\
4 &  0.112722 & 0.112134 & 0.111551 & 0.110973 & 0.110399 & 0.109831 \\
5 &  0.055715 & 0.055375 & 0.055038 & 0.054703 & 0.054371 & 0.054041 \\
\hline
&&&&&&\\
&&&&&&\\
\hline
rate  & Algo 1  & Algo 1  & Algo 1  & Algo 1  & Algo 1  & Algo 1  \\
      & $\beta =$ 33 & $\beta =$ 34 & $\beta =$ 35 & $\beta =$ 36 & $\beta =$ 37 & $\beta =$ 38 \\
\hline
2 &  unstable & unstable & unstable & unstable & unstable & unstable \\
3 &  0.211198 & 0.210376 & 0.209568 & 0.208775 & 0.207996 & 0.207233 \\
4 &  0.109268 & 0.108710 & 0.108157 & 0.107610 & 0.107069 & 0.106536 \\
5 &  0.053714 & 0.053389 & 0.053068 & 0.052749 & 0.052433 & 0.052121 \\
\hline
&&&&&&\\
&&&&&&\\
\hline
rate  & Algo 1  & Algo 1  & Algo 1  & Algo 1  & Algo 1  & Algo 1  \\
      & $\beta =$ 39 & $\beta =$ 40 & $\beta =$ 41 & $\beta =$ 42 & $\beta =$ 43 & $\beta =$ 44 \\
\hline
2 &  unstable & unstable & unstable & unstable & unstable & unstable \\
3 &  0.206485 & 0.205754 & 0.205038 & 0.204341 & 0.203850 & unstable \\
4 &  0.106004 & 0.105481 & 0.104966 & 0.104455 & 0.103949 & 0.103452 \\
5 &  0.051811 & 0.051504 & 0.051201 & 0.050901 & 0.050604 & 0.050310 \\
\hline
\clearpage
&\\
&\\
\hline
rate  & Algo 1  \\
      & $\beta =$ 45 \\
\hline
2 &  unstable \\
3 &  unstable \\
4 &  0.102961 \\
5 &  0.050021 \\
\hline
\hline
\end{longtable}

%% file: tables/prj_prs-coret_v3_m2.tex
\begin{longtable}{lllllll }
\caption{Numerical results of Fernandez fully decoupled  scheme (Fern FD) and Algorithm 2 (Algo 2)}
\label{table:prj_prs-coret_v3_m2}\\
\hline
\hline
rate &  Fern FD  & Algo 2  & Algo 2  & Algo 2  & Algo 2  & Algo 2  \\
     &                 &  $\theta =$ 10 & $\theta =$ 11 & $\theta =$ 12 & $\theta =$ 13 & $\theta =$ 14 \\
\hline
2 &  0.437713 & 0.420421 & 0.418264 & 0.416110 & 0.413961 & 0.411817 \\
3 &  0.243562 & 0.231346 & 0.229813 & 0.228279 & 0.226744 & 0.225208 \\
4 &  0.129731 & 0.123637 & 0.122873 & 0.122109 & 0.121345 & 0.120580 \\
5 &  0.065497 & 0.063052 & 0.062746 & 0.062441 & 0.062136 & 0.061830 \\
\hline
&&&&&&\\
&&&&&&\\
\hline
rate  & Algo 2  & Algo 2  & Algo 2  & Algo 2  & Algo 2  & Algo 2  \\
      & $\theta =$ 15 & $\theta =$ 16 & $\theta =$ 17 & $\theta =$ 18 & $\theta =$ 19 & $\theta =$ 20 \\
\hline
2 &  0.409654 & unstable & unstable & unstable & unstable & unstable \\
3 &  0.223672 & 0.222135 & 0.220599 & 0.219063 & 0.217527 & 0.215992 \\
4 &  0.119816 & 0.119050 & 0.118285 & 0.117519 & 0.116754 & 0.115988 \\
5 &  0.061525 & 0.061220 & 0.060915 & 0.060610 & 0.060305 & 0.060000 \\
\hline
&&&&&&\\
&&&&&&\\
\hline
rate  & Algo 2  & Algo 2  & Algo 2  & Algo 2  & Algo 2  & Algo 2  \\
      & $\theta =$ 21 & $\theta =$ 22 & $\theta =$ 23 & $\theta =$ 24 & $\theta =$ 25 & $\theta =$ 26 \\
\hline
2 &  unstable & unstable & unstable & unstable & unstable & unstable \\
3 &  0.214458 & 0.212924 & 0.211393 & 0.209862 & 0.208879 & unstable \\
4 &  0.115222 & 0.114456 & 0.113690 & 0.112924 & 0.112158 & 0.111392 \\
5 &  0.059695 & 0.059391 & 0.059086 & 0.058782 & 0.058477 & 0.058173 \\
\hline
\clearpage
&&&&&&\\
&&&&&&\\
\hline
rate  & Algo 2  & Algo 2  & Algo 2  & Algo 2  & Algo 2  & Algo 2  \\
      & $\theta =$ 27 & $\theta =$ 28 & $\theta =$ 29 & $\theta =$ 30 & $\theta =$ 31 & $\theta =$ 32 \\
\hline
2 &  unstable & unstable & unstable & unstable & unstable & unstable \\
3 &  unstable & unstable & unstable & unstable & unstable & unstable \\
4 &  0.110627 & 0.109861 & 0.109096 & 0.108330 & 0.107566 & 0.106801 \\
5 &  0.057869 & 0.057565 & 0.057261 & 0.056958 & 0.056654 & 0.056351 \\
\hline
&&&&&&\\
&&&&&&\\
\hline
rate  & Algo 2  & Algo 2  & Algo 2  & Algo 2  & Algo 2  & Algo 2  \\
      & $\theta =$ 33 & $\theta =$ 34 & $\theta =$ 35 & $\theta =$ 36 & $\theta =$ 37 & $\theta =$ 38 \\
\hline
2 &  unstable & unstable & unstable & unstable & unstable & unstable \\
3 &  unstable & unstable & unstable & unstable & unstable & unstable \\
4 &  0.106036 & 0.105272 & 0.104509 & 0.103746 & 0.102984 & 0.102222 \\
5 &  0.056048 & 0.055745 & 0.055442 & 0.055139 & 0.054837 & 0.054534 \\
\hline
&&&&&&\\
&&&&&&\\
\hline
rate  & Algo 2  & Algo 2  & Algo 2  & Algo 2  & Algo 2  & Algo 2  \\
      & $\theta =$ 39 & $\theta =$ 40 & $\theta =$ 41 & $\theta =$ 42 & $\theta =$ 43 & $\theta =$ 44 \\
\hline
2 &  unstable & unstable & unstable & unstable & unstable & unstable \\
3 &  unstable & unstable & unstable & unstable & unstable & unstable \\
4 &  0.101463 & 0.100700 & 0.099940 & 0.099186 & unstable & unstable \\
5 &  0.054232 & 0.053931 & 0.053629 & 0.053327 & 0.053027 & 0.052726 \\
\hline
\clearpage
&\\
&\\
\hline
rate  & Algo 2  \\
      & $\theta =$ 45 \\
\hline
2 &  unstable \\
3 &  unstable \\
4 &  unstable \\
5 &  0.052425 \\
\hline
\hline
\end{longtable}

%% file: tables/prj_prs-coret_Van-Kan_m3.tex
\begin{longtable}{lllllll }
\caption{Numerical results of Fernandez fully decoupled  scheme (Fern FD) and Algorithm 3 (Algo 3)}
\label{table:prj_prs-coret_Van-Kan_m3}\\
\hline
\hline
rate &  Fern FD  & Algo 3  & Algo 3  & Algo 3  & Algo 3  & Algo 3  \\
     &                 &  $\xi =$ 10 & $\xi =$ 11 & $\xi =$ 12 & $\xi =$ 13 & $\xi =$ 14 \\
\hline
2 &  0.437713 & 0.413443 & 0.408376 & 0.403425 & 0.398606 & 0.393939 \\
3 &  0.243562 & 0.248517 & 0.243651 & 0.238858 & 0.234148 & 0.229535 \\
4 &  0.129731 & 0.136471 & 0.132807 & 0.129173 & 0.125574 & 0.122016 \\
5 &  0.065497 & 0.069889 & 0.067621 & 0.065367 & 0.063132 & 0.060917 \\
\hline
&&&&&&\\
&&&&&&\\
\hline
rate  & Algo 3  & Algo 3  & Algo 3  & Algo 3  & Algo 3  & Algo 3  \\
      & $\xi =$ 15 & $\xi =$ 16 & $\xi =$ 17 & $\xi =$ 18 & $\xi =$ 19 & $\xi =$ 20 \\
\hline
2 &  0.389442 & 0.385136 & 0.381043 & 0.377186 & 0.373591 & 0.370281 \\
3 &  0.225031 & 0.220650 & 0.216408 & 0.212321 & 0.208408 & 0.204686 \\
4 &  0.118508 & 0.115057 & 0.111672 & 0.108363 & 0.105142 & 0.102021 \\
5 &  0.058728 & 0.056566 & 0.054439 & 0.052351 & 0.050309 & 0.048320 \\
\hline
&&&&&&\\
&&&&&&\\
\hline
rate  & Algo 3  & Algo 3  & Algo 3  & Algo 3  & Algo 3  & Algo 3  \\
      & $\xi =$ 21 & $\xi =$ 22 & $\xi =$ 23 & $\xi =$ 24 & $\xi =$ 25 & $\xi =$ 26 \\
\hline
2 &  0.367286 & 0.364632 & 0.362348 & 0.360462 & 0.359005 & 0.358005 \\
3 &  0.201178 & 0.197904 & 0.194886 & 0.192147 & 0.189712 & 0.187604 \\
4 &  0.099015 & 0.096139 & 0.093410 & 0.090847 & 0.088472 & 0.086299 \\
5 &  0.046394 & 0.044540 & 0.042769 & 0.041095 & 0.039531 & 0.038095 \\
\hline
\clearpage
&&&&&&\\
&&&&&&\\
\hline
rate  & Algo 3  & Algo 3  & Algo 3  & Algo 3  & Algo 3  & Algo 3  \\
      & $\xi =$ 27 & $\xi =$ 28 & $\xi =$ 29 & $\xi =$ 30 & $\xi =$ 31 & $\xi =$ 32 \\
\hline
2 &  0.357492 & 0.357517 & 0.358058 & 0.359165 & 0.360862 & 0.363169 \\
3 &  0.185846 & 0.184461 & 0.183471 & 0.182897 & 0.182757 & 0.183065 \\
4 &  0.084358 & 0.082669 & 0.081254 & 0.080136 & 0.079329 & 0.078857 \\
5 &  0.036802 & 0.035672 & 0.034723 & 0.033973 & 0.033437 & 0.033131 \\
\hline
&&&&&&\\
&&&&&&\\
\hline
rate  & Algo 3  & Algo 3  & Algo 3  & Algo 3  & Algo 3  & Algo 3  \\
      & $\xi =$ 33 & $\xi =$ 34 & $\xi =$ 35 & $\xi =$ 36 & $\xi =$ 37 & $\xi =$ 38 \\
\hline
2 &  0.366107 & 0.369690 & 0.373933 & 0.378845 & 0.384418 & 0.392112 \\
3 &  0.183835 & 0.185076 & 0.186796 & 0.188994 & 0.191672 & 0.194826 \\
4 &  0.078730 & 0.078957 & 0.079543 & 0.080489 & 0.081789 & 0.083434 \\
5 &  0.033061 & 0.033234 & 0.033648 & 0.034298 & 0.035172 & 0.036259 \\
\hline
&&&&&&\\
&&&&&&\\
\hline
rate  & Algo 3  & Algo 3  & Algo 3  & Algo 3  & Algo 3  & Algo 3  \\
      & $\xi =$ 39 & $\xi =$ 40 & $\xi =$ 41 & $\xi =$ 42 & $\xi =$ 43 & $\xi =$ 44 \\
\hline
2 &  unstable & unstable & unstable & unstable & unstable & unstable \\
3 &  0.198449 & 0.202535 & 0.207070 & 0.212044 & 0.217443 & 0.223253 \\
4 &  0.085413 & 0.087711 & 0.090310 & 0.093194 & 0.096343 & 0.099741 \\
5 &  0.037542 & 0.039004 & 0.040630 & 0.042402 & 0.044305 & 0.046326 \\
\hline
\clearpage
&\\
&\\
\hline
rate  & Algo 3  \\
      & $\xi =$ 45 \\
\hline
2 &  unstable \\
3 &  0.229459 \\
4 &  0.103370 \\
5 &  0.048453 \\
\hline
\hline
\end{longtable}

%% file: tables/rn_m1_selected.tex
\begin{longtable}{lllll }
\caption{Numerical results of Fernandez Explicit Robin-Neumann scheme (Fern ERN) and Algorithm 1 (Algo 1) at selected $ \beta$}
\label{table:rn_m1_selected}\\
\hline
\hline
rate &  Fern ERN & Algo 1 & \multicolumn{2}{l}{decrement of errors} \\
     &           &  $\beta = 43 $ & values  & percents(\%) \\
\hline
2 &  0.435176 & unstable & N/A & N/A  \\
3 &  0.241766 & 0.203850 & 0.037916 & 15.6829  \\
4 &  0.128616 & 0.103949 & 0.024667 & 19.1788  \\
5 &  0.064847 & 0.050604 & 0.014243 & 21.9640  \\
\hline
\hline
\end{longtable}

%% file: tables/prj_prs-coret_v3_m2_selected.tex
\begin{longtable}{lllll }
\caption{Numerical results of Fernandez fully decoupled  scheme (Fern FD) and Algorithm 2 (Algo 2) at selected $ \theta$}
\label{table:prj_prs-coret_v3_m2_selected}\\
\hline
\hline
rate &  Fern FD & Algo 2 & \multicolumn{2}{l}{decrement of errors} \\
     &           &  $\theta = 25 $ & values  & percents(\%) \\
\hline
2 &  0.437713 & unstable & N/A & N/A  \\
3 &  0.243562 & 0.208879 & 0.034683 & 14.2399  \\
4 &  0.129731 & 0.112158 & 0.017573 & 13.5457  \\
5 &  0.065497 & 0.058477 & 0.00702 & 10.7180  \\
\hline
\hline
\end{longtable}

%% file: tables/prj_prs-coret_Van-Kan_m3_selected.tex
\begin{longtable}{lllll }
\caption{Numerical results of Fernandez fully decoupled  scheme (Fern FD) and Algorithm 3 (Algo 3) at selected $ \xi$}
\label{table:prj_prs-coret_Van-Kan_m3_selected}\\
\hline
\hline
rate &  Fern FD & Algo 3 & \multicolumn{2}{l}{decrement of errors} \\
     &           &  $\xi = 31 $ & values  & percents(\%) \\
\hline
2 &  0.437713 & 0.360862 & 0.076851 & 17.5574  \\
3 &  0.243562 & 0.182757 & 0.060805 & 24.9649  \\
4 &  0.129731 & 0.079329 & 0.050402 & 38.8512  \\
5 &  0.065497 & 0.033437 & 0.03206 & 48.9488  \\
\hline
& & & &  \\
\hline
rate &  Fern FD & Algo 3 & \multicolumn{2}{l}{decrement of errors} \\
     &           &  $\xi = 32 $ & values  & percents(\%) \\
\hline
2 &  0.437713 & 0.363169 & 0.074544 & 17.0303  \\
3 &  0.243562 & 0.183065 & 0.060497 & 24.8384  \\
4 &  0.129731 & 0.078857 & 0.050874 & 39.2150  \\
5 &  0.065497 & 0.033131 & 0.032366 & 49.4160  \\
\hline
& & & &  \\
\hline
rate &  Fern FD & Algo 3 & \multicolumn{2}{l}{decrement of errors} \\
     &           &  $\xi = 33 $ & values  & percents(\%) \\
\hline
2 &  0.437713 & 0.366107 & 0.071606 & 16.3591  \\
3 &  0.243562 & 0.183835 & 0.059727 & 24.5223  \\
4 &  0.129731 & 0.078730 & 0.051001 & 39.3129  \\
5 &  0.065497 & 0.033061 & 0.032436 & 49.5229  \\
\hline
\hline
\end{longtable}

%% file: images/rn_m1_r1_beta43.tex
\begin{figure}[h]
  \centering
\begin{subfigure}[h]{0.49\textwidth}
    \centering
    \includegraphics[width=\textwidth]{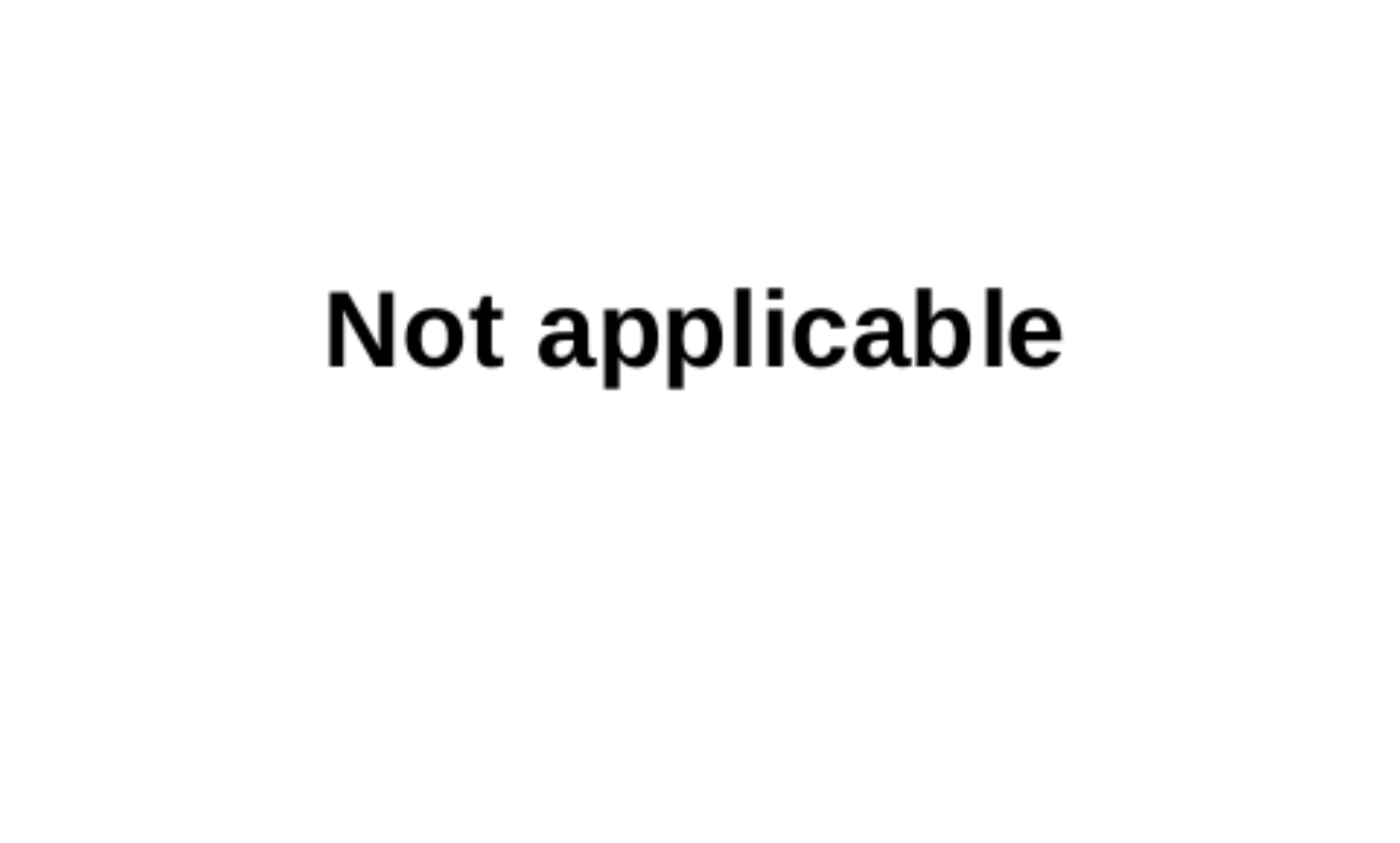}
    \caption{rate = 2}
    \label{fig:rn_m1_r1_rate2_beta43}
  \end{subfigure}
\begin{subfigure}[h]{0.49\textwidth}
    \centering
    \includegraphics[width=\textwidth]{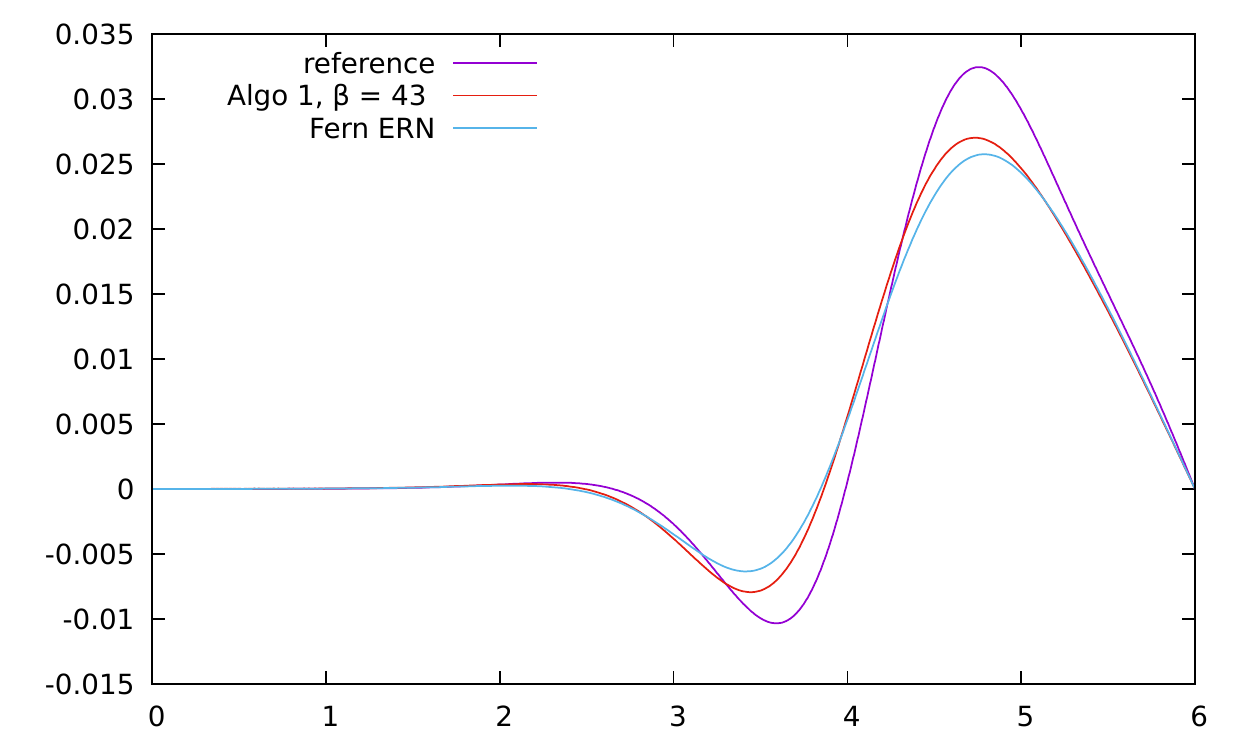}
    \caption{rate = 3}
    \label{fig:rn_m1_r1_rate3_beta43}
  \end{subfigure}

\begin{subfigure}[h]{0.49\textwidth}
    \centering
    \includegraphics[width=\textwidth]{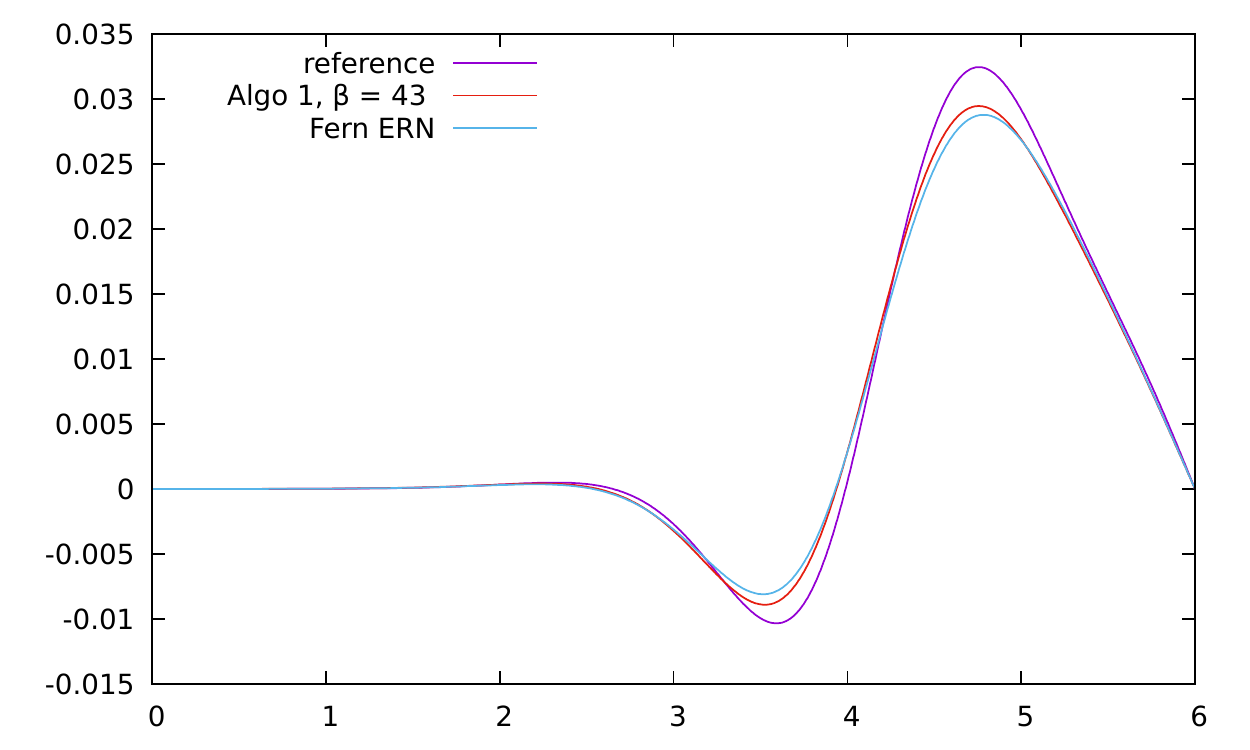}
    \caption{rate = 4}
    \label{fig:rn_m1_r1_rate4_beta43}
  \end{subfigure}
\begin{subfigure}[h]{0.49\textwidth}
    \centering
    \includegraphics[width=\textwidth]{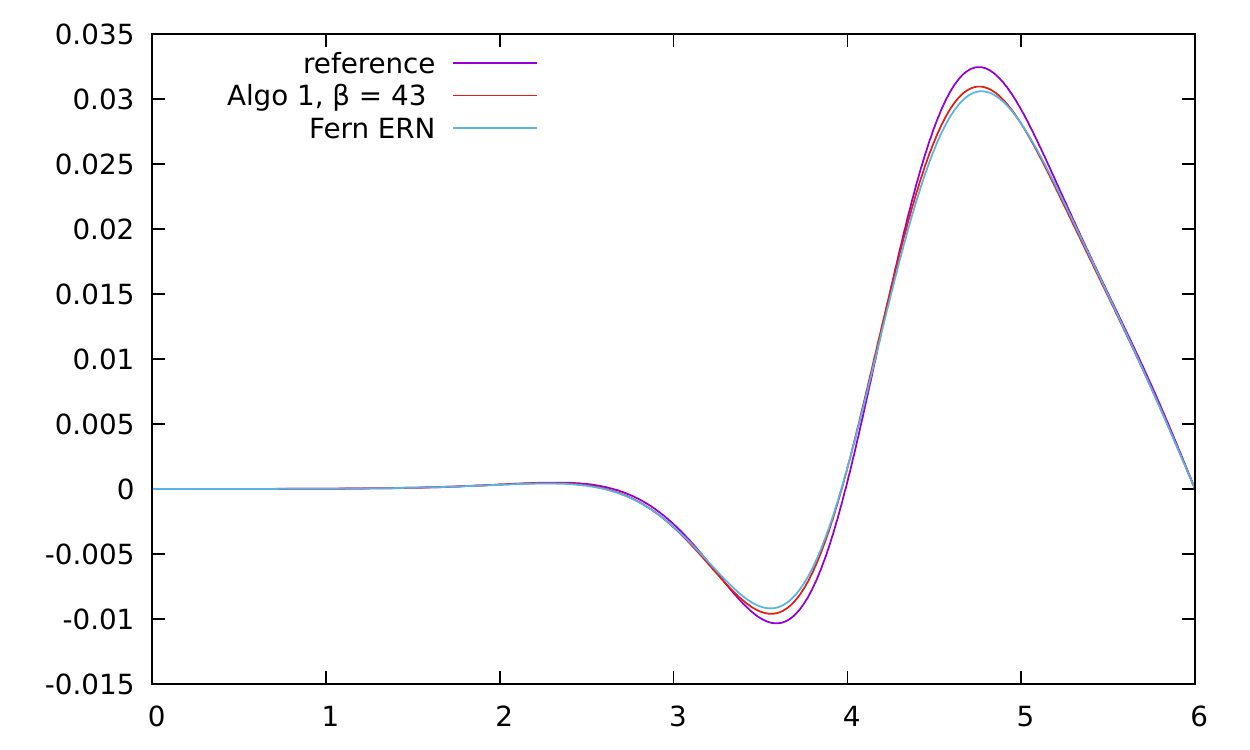}
    \caption{rate = 5}
    \label{fig:rn_m1_r1_rate5_beta43}
  \end{subfigure}
  \caption{Structure displacement of Fernandez Explicit Robin-Neumann scheme  (Fern ERN) and Algorithm 1 (Algo 1) with $ \beta = 43 $ at final time}
  \label{fig:rn_m1_r1_beta43}
\end{figure}

%% file: images/prj_prs-coret_v3_m2_r1_beta25.tex
\begin{figure}[h]
  \centering
\begin{subfigure}[h]{0.49\textwidth}
    \centering
    \includegraphics[width=\textwidth]{not_applicable.pdf}
    \caption{rate = 2}
    \label{fig:prj_prs-coret_v3_m2_r1_rate2_beta25}
  \end{subfigure}
\begin{subfigure}[h]{0.49\textwidth}
    \centering
    \includegraphics[width=\textwidth]{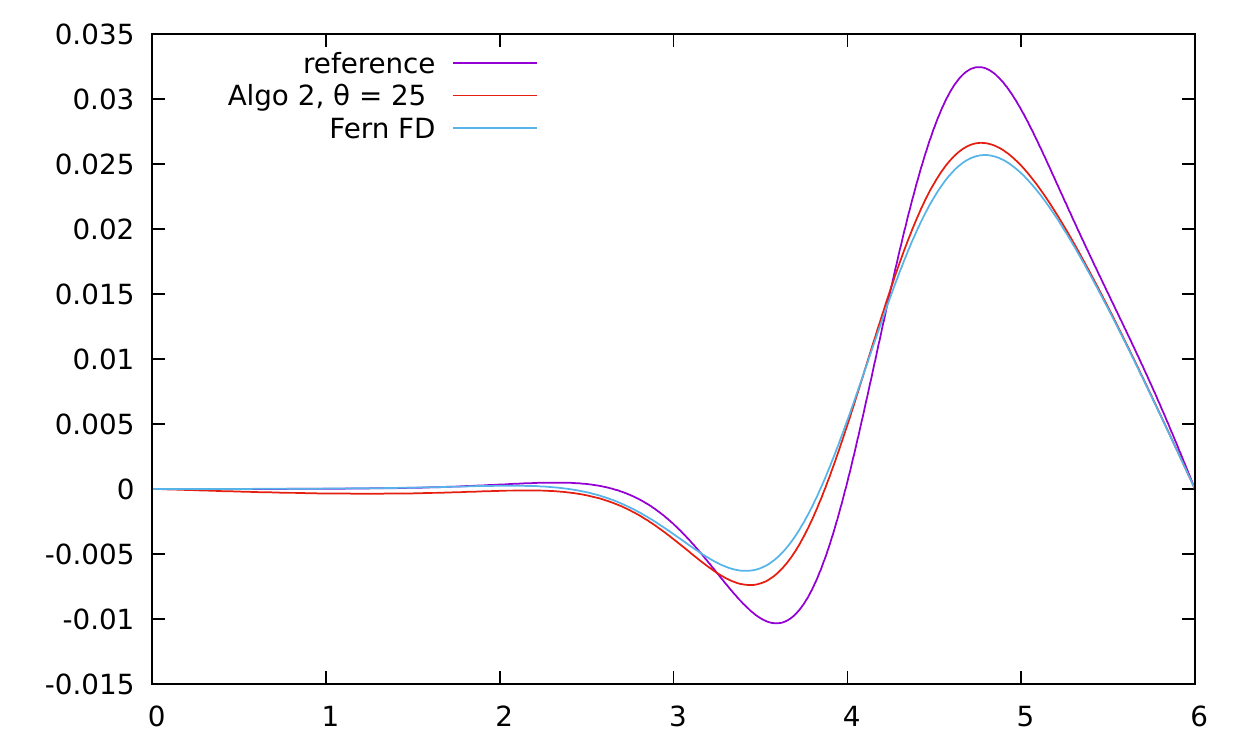}
    \caption{rate = 3}
    \label{fig:prj_prs-coret_v3_m2_r1_rate3_beta25}
  \end{subfigure}

\begin{subfigure}[h]{0.49\textwidth}
    \centering
    \includegraphics[width=\textwidth]{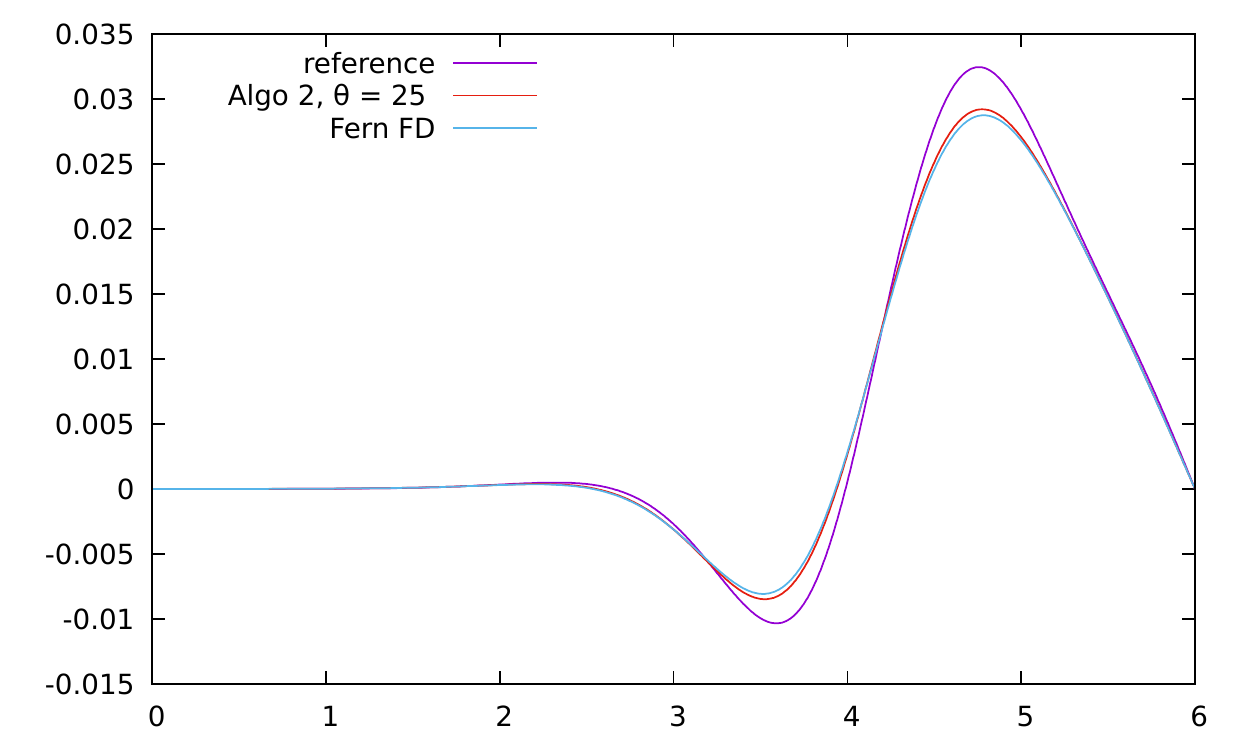}
    \caption{rate = 4}
    \label{fig:prj_prs-coret_v3_m2_r1_rate4_beta25}
  \end{subfigure}
\begin{subfigure}[h]{0.49\textwidth}
    \centering
    \includegraphics[width=\textwidth]{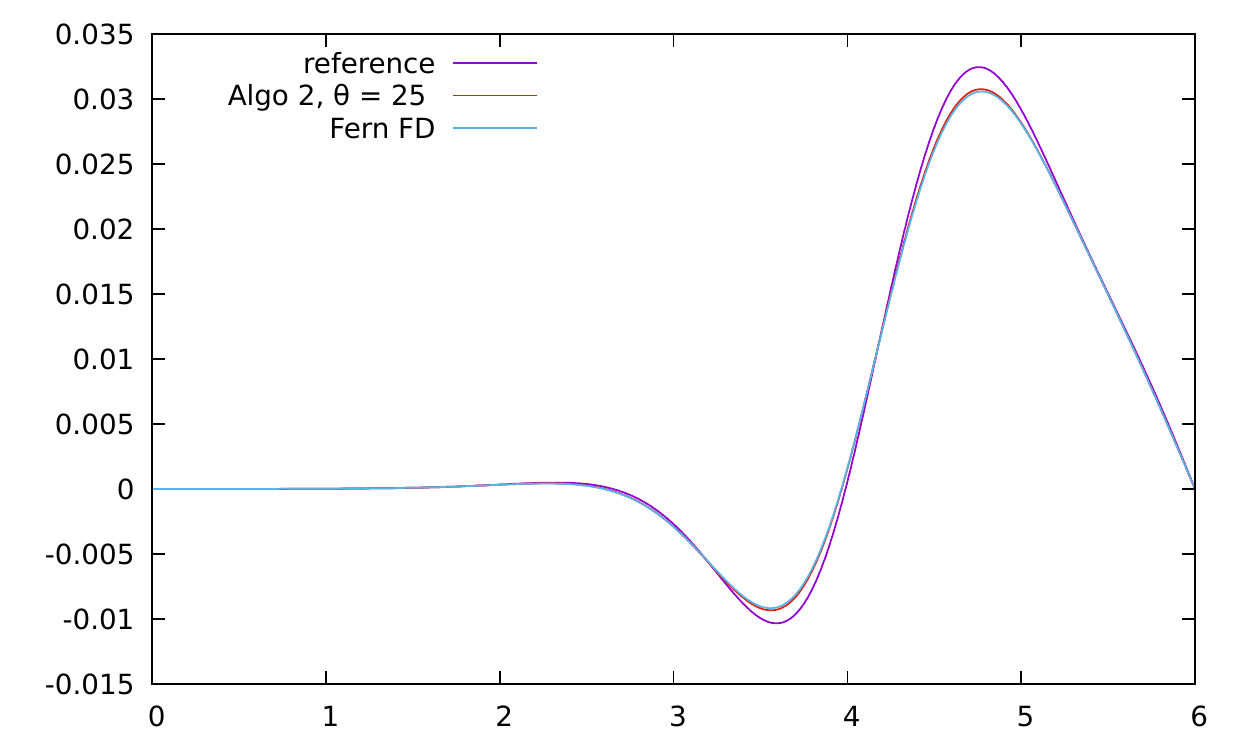}
    \caption{rate = 5}
    \label{fig:prj_prs-coret_v3_m2_r1_rate5_beta25}
  \end{subfigure}
  \caption{Structure displacement of Fernandez fully decoupled  scheme  (Fern FD) and Algorithm 2 (Algo 2) with $ \theta = 25 $ at final time}
  \label{fig:prj_prs-coret_v3_m2_r1_beta25}
\end{figure}

%% file: images/prj_prs-coret_Van-Kan_m3_r0_beta31.tex
\begin{figure}[h]
  \centering
\begin{subfigure}[h]{0.49\textwidth}
    \centering
    \includegraphics[width=\textwidth]{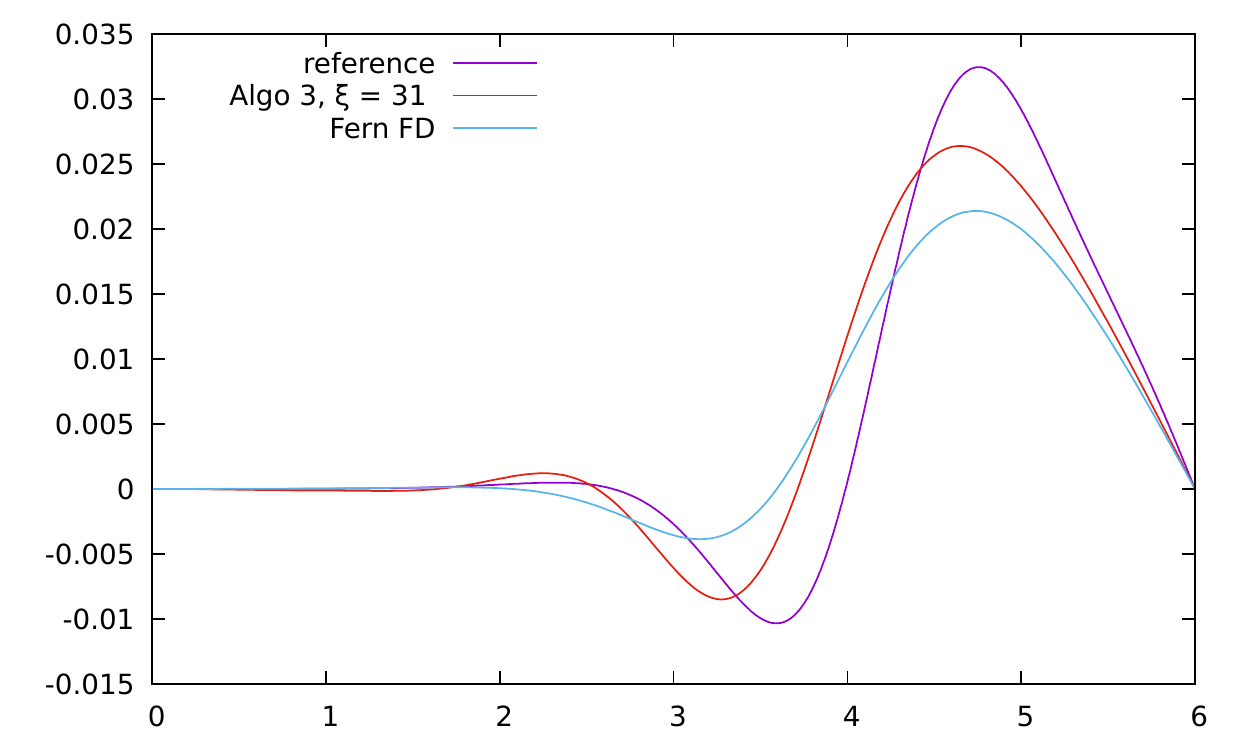}
    \caption{rate = 2}
    \label{fig:prj_prs-coret_Van-Kan_m3_r0_rate2_beta31}
  \end{subfigure}
\begin{subfigure}[h]{0.49\textwidth}
    \centering
    \includegraphics[width=\textwidth]{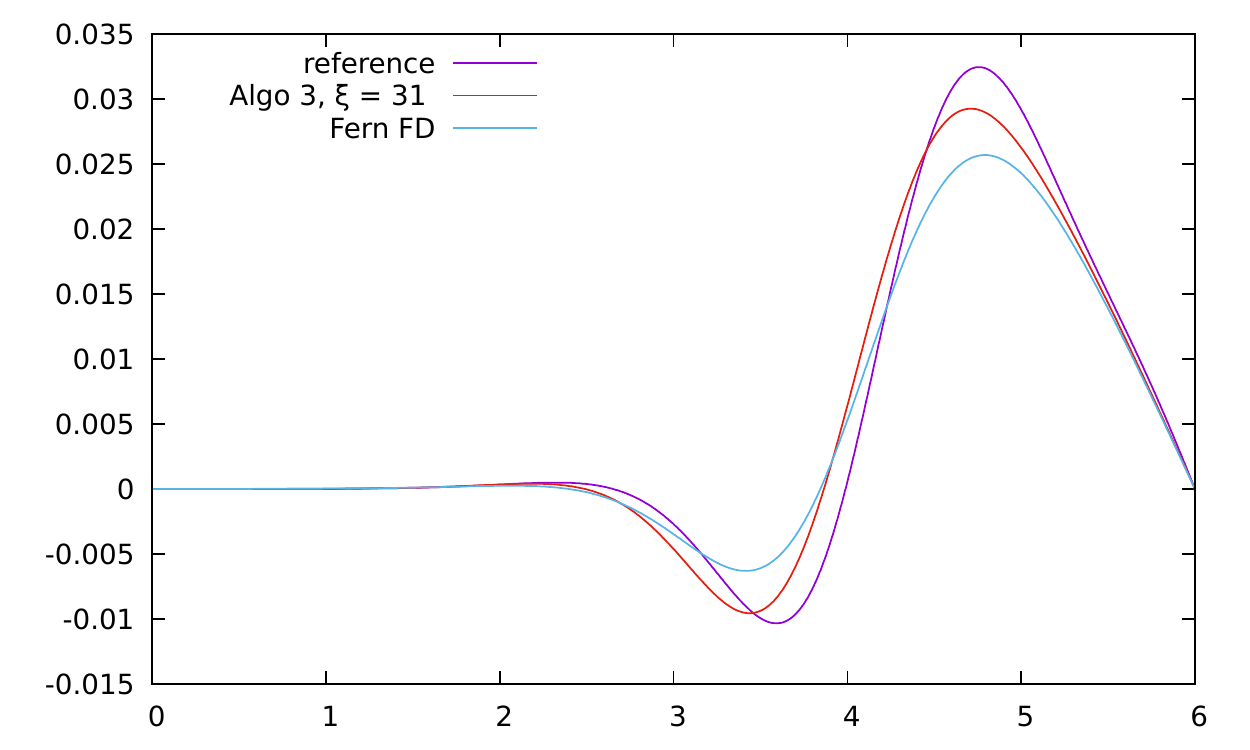}
    \caption{rate = 3}
    \label{fig:prj_prs-coret_Van-Kan_m3_r0_rate3_beta31}
  \end{subfigure}

\begin{subfigure}[h]{0.49\textwidth}
    \centering
    \includegraphics[width=\textwidth]{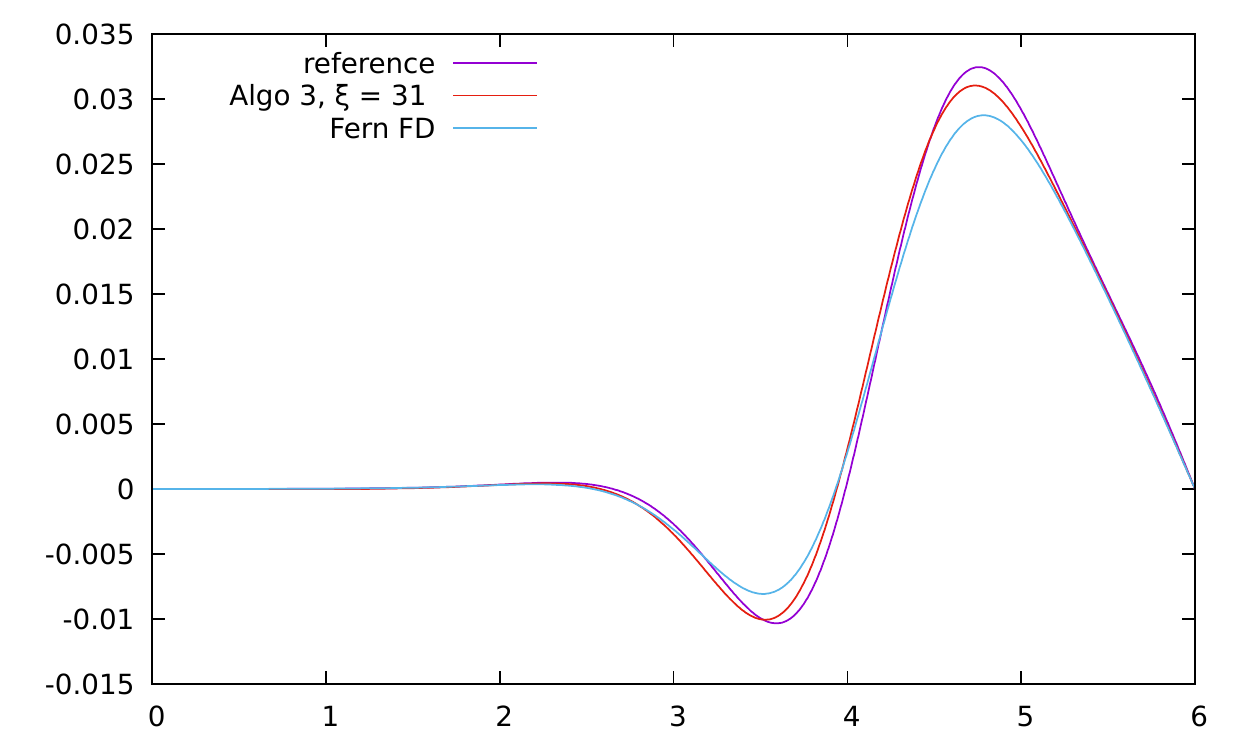}
    \caption{rate = 4}
    \label{fig:prj_prs-coret_Van-Kan_m3_r0_rate4_beta31}
  \end{subfigure}
\begin{subfigure}[h]{0.49\textwidth}
    \centering
    \includegraphics[width=\textwidth]{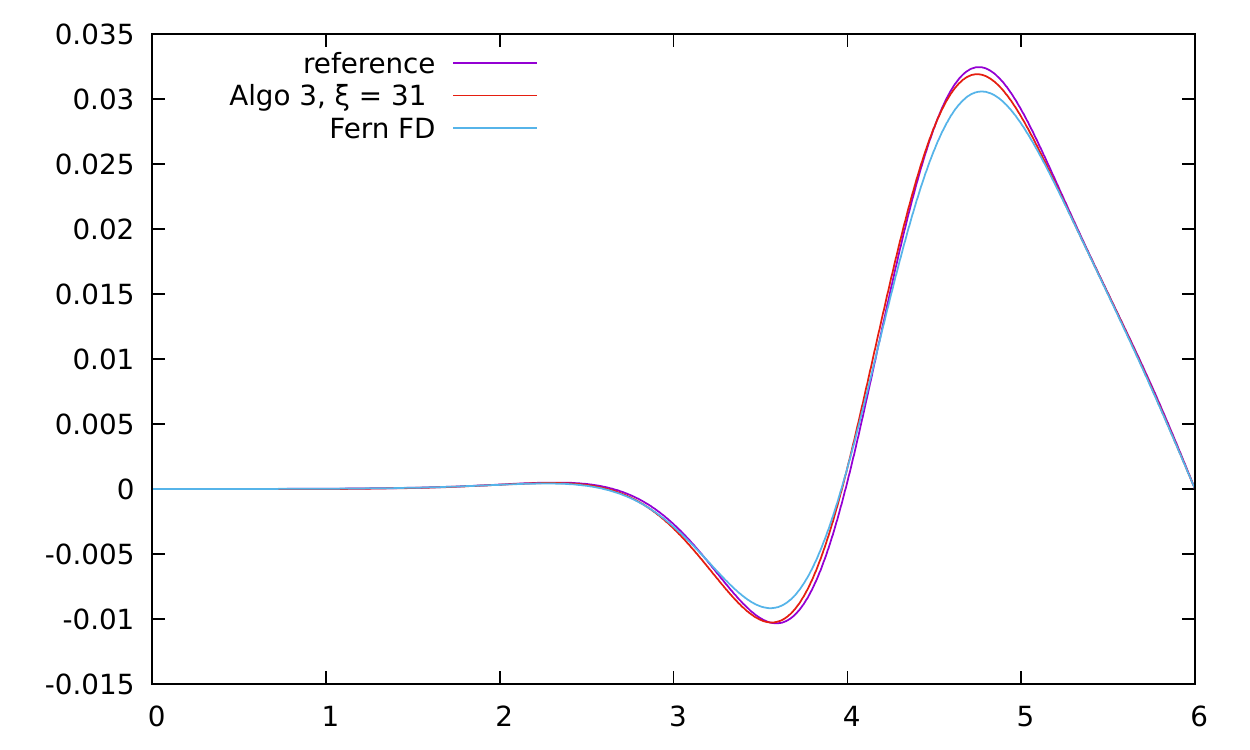}
    \caption{rate = 5}
    \label{fig:prj_prs-coret_Van-Kan_m3_r0_rate5_beta31}
  \end{subfigure}
  \caption{Structure displacement of Fernandez fully decoupled  scheme  (Fern FD) and Algorithm 3 (Algo 3) with $ \xi = 31 $ at final time}
  \label{fig:prj_prs-coret_Van-Kan_m3_r0_beta31}
\end{figure}

%% file: images/prj_prs-coret_Van-Kan_m3_r0_beta32.tex
\begin{figure}[h]
  \centering
\begin{subfigure}[h]{0.49\textwidth}
    \centering
    \includegraphics[width=\textwidth]{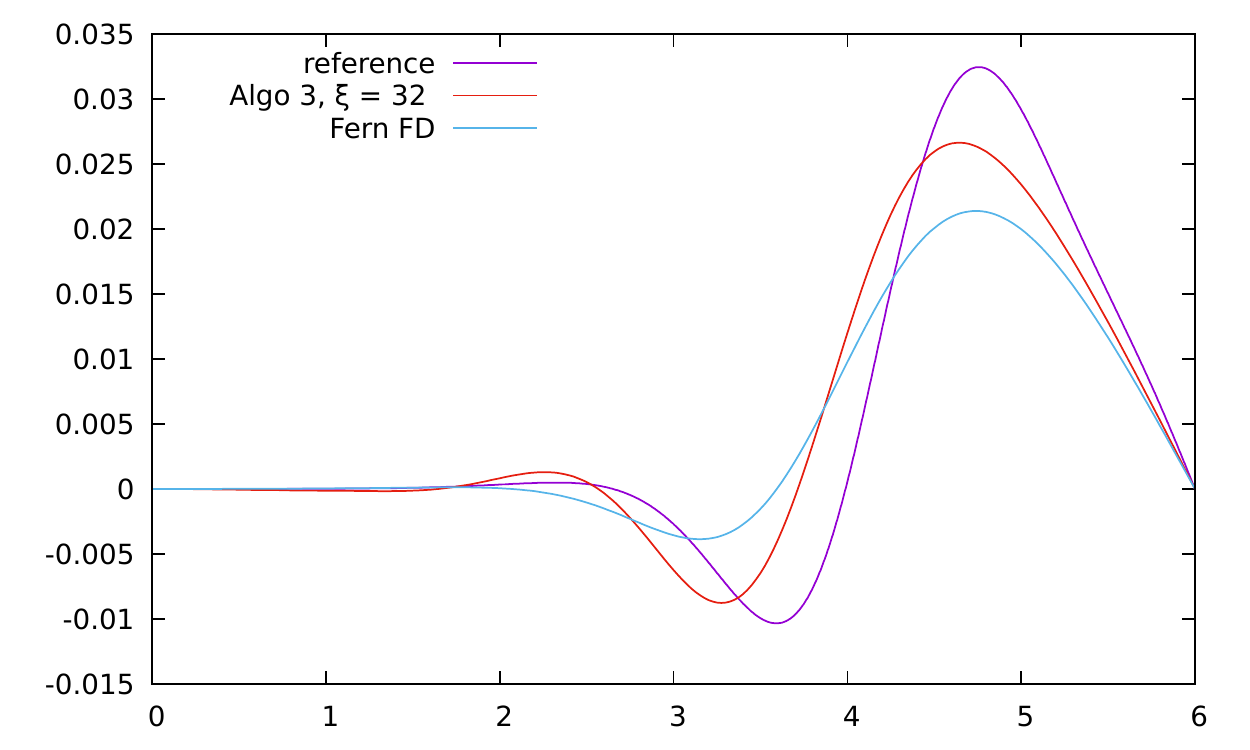}
    \caption{rate = 2}
    \label{fig:prj_prs-coret_Van-Kan_m3_r0_rate2_beta32}
  \end{subfigure}
\begin{subfigure}[h]{0.49\textwidth}
    \centering
    \includegraphics[width=\textwidth]{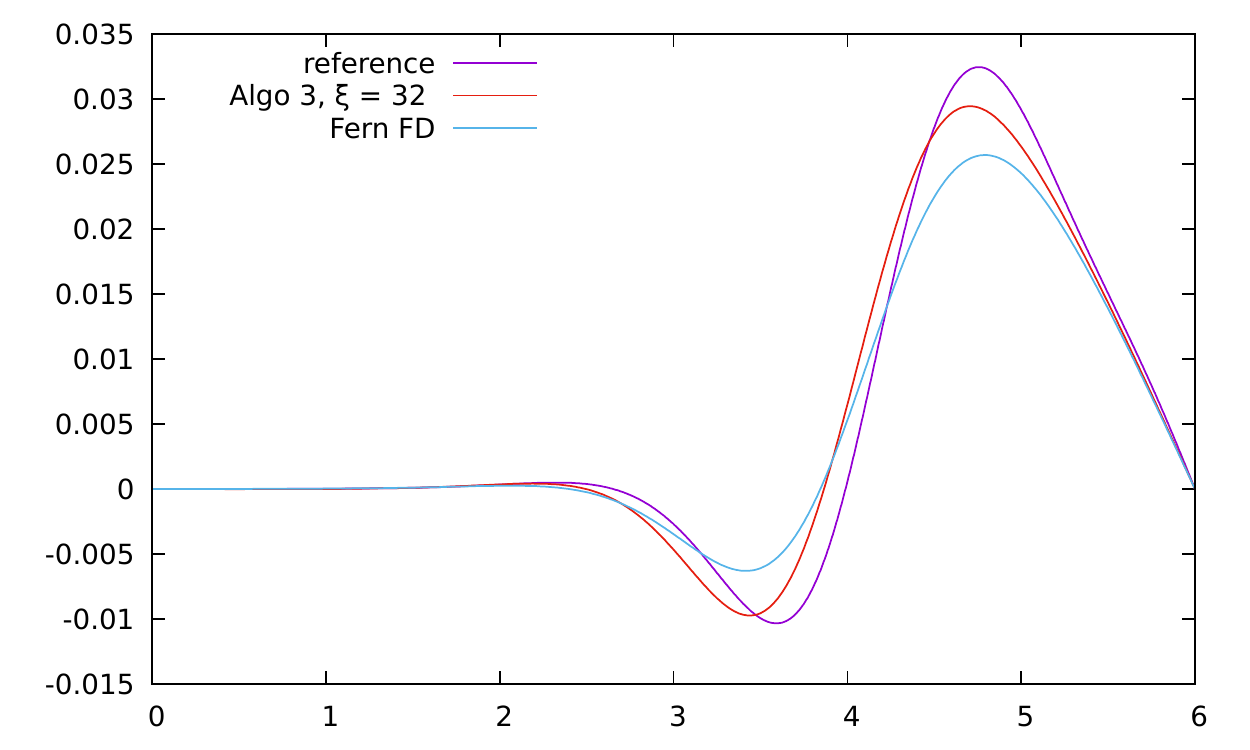}
    \caption{rate = 3}
    \label{fig:prj_prs-coret_Van-Kan_m3_r0_rate3_beta32}
  \end{subfigure}

\begin{subfigure}[h]{0.49\textwidth}
    \centering
    \includegraphics[width=\textwidth]{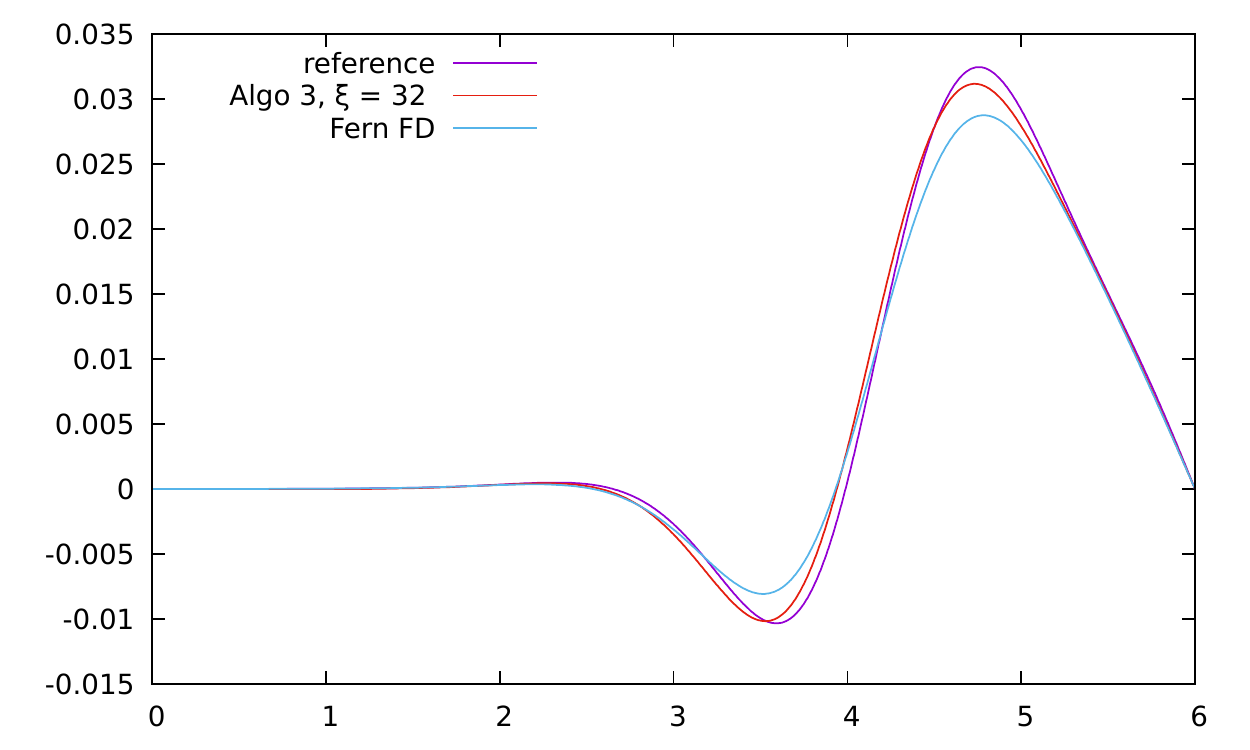}
    \caption{rate = 4}
    \label{fig:prj_prs-coret_Van-Kan_m3_r0_rate4_beta32}
  \end{subfigure}
\begin{subfigure}[h]{0.49\textwidth}
    \centering
    \includegraphics[width=\textwidth]{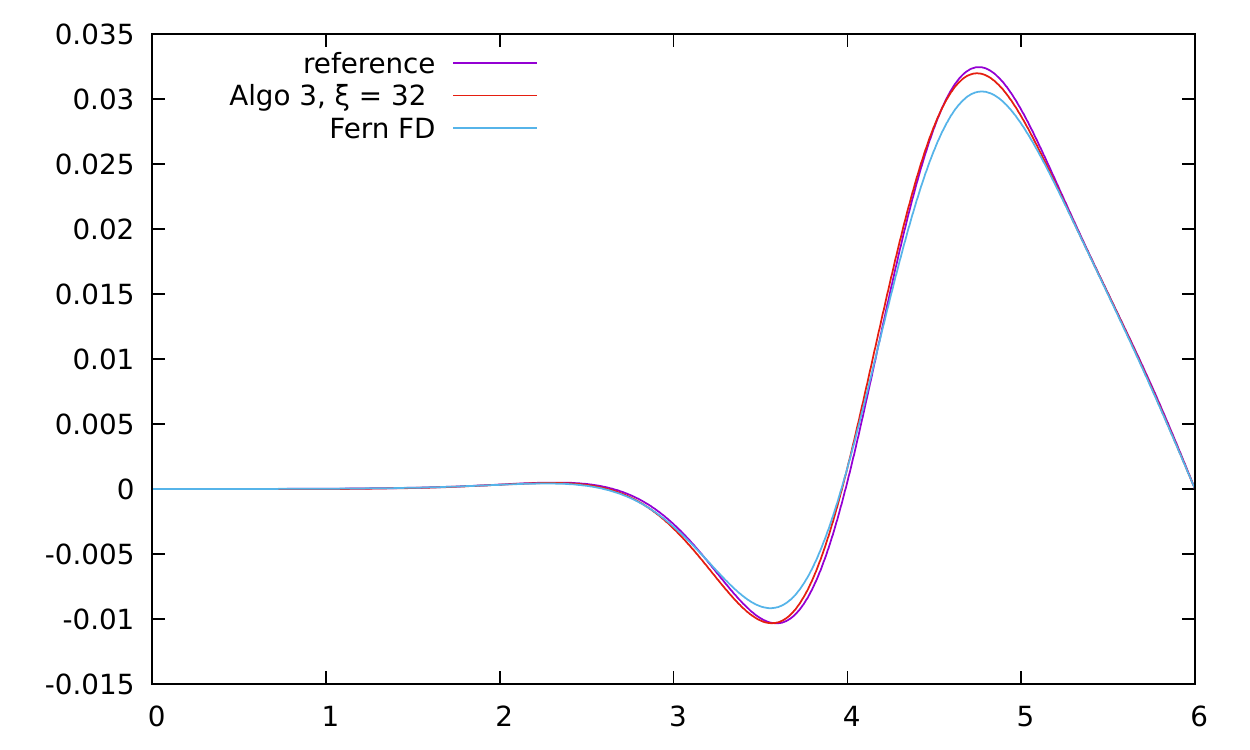}
    \caption{rate = 5}
    \label{fig:prj_prs-coret_Van-Kan_m3_r0_rate5_beta32}
  \end{subfigure}
  \caption{Structure displacement of Fernandez fully decoupled  scheme  (Fern FD) and Algorithm 3 (Algo 3) with $ \xi = 32 $ at final time}
  \label{fig:prj_prs-coret_Van-Kan_m3_r0_beta32}
\end{figure}

%% file: images/prj_prs-coret_Van-Kan_m3_r0_beta33.tex
\begin{figure}[h]
  \centering
\begin{subfigure}[h]{0.49\textwidth}
    \centering
    \includegraphics[width=\textwidth]{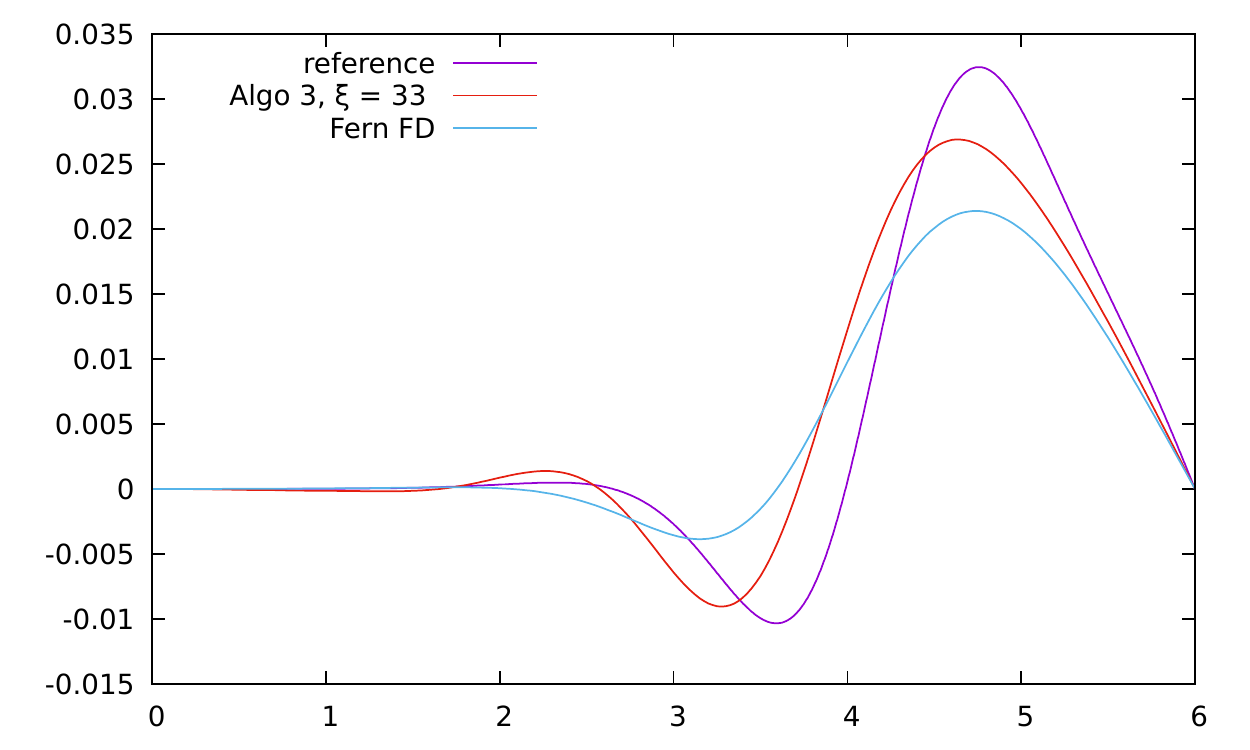}
    \caption{rate = 2}
    \label{fig:prj_prs-coret_Van-Kan_m3_r0_rate2_beta33}
  \end{subfigure}
\begin{subfigure}[h]{0.49\textwidth}
    \centering
    \includegraphics[width=\textwidth]{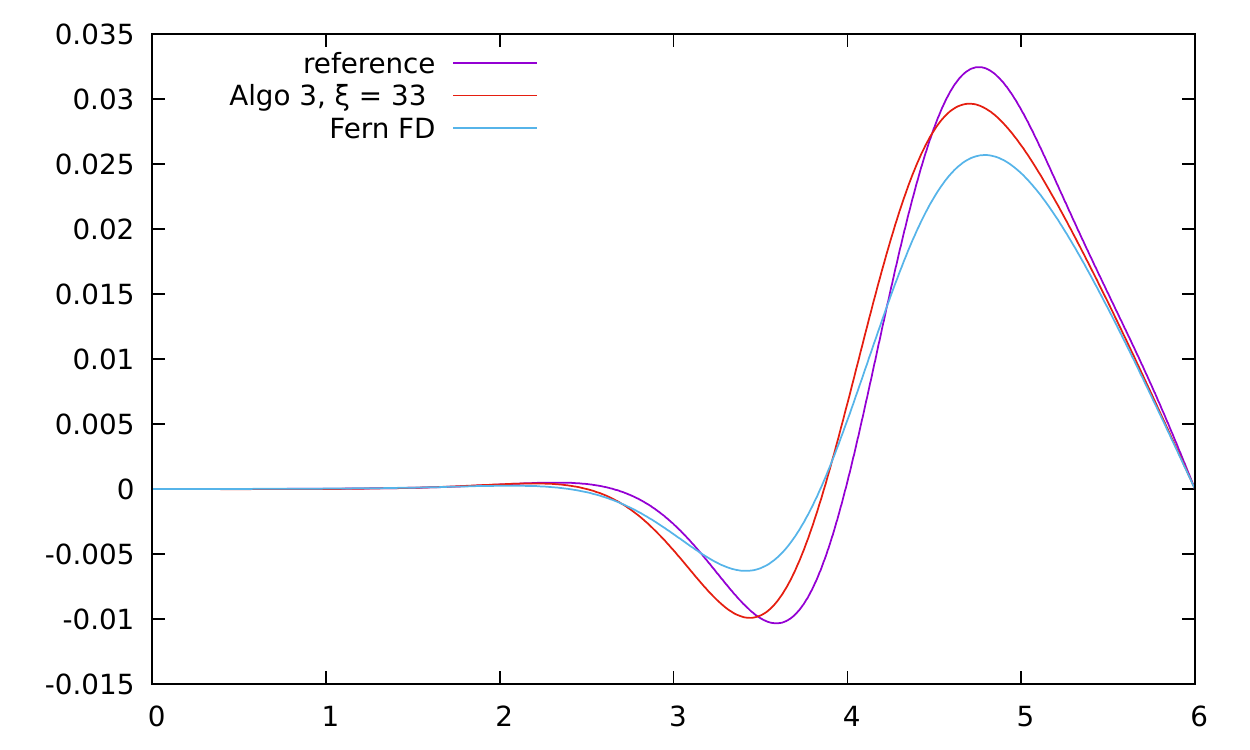}
    \caption{rate = 3}
    \label{fig:prj_prs-coret_Van-Kan_m3_r0_rate3_beta33}
  \end{subfigure}

\begin{subfigure}[h]{0.49\textwidth}
    \centering
    \includegraphics[width=\textwidth]{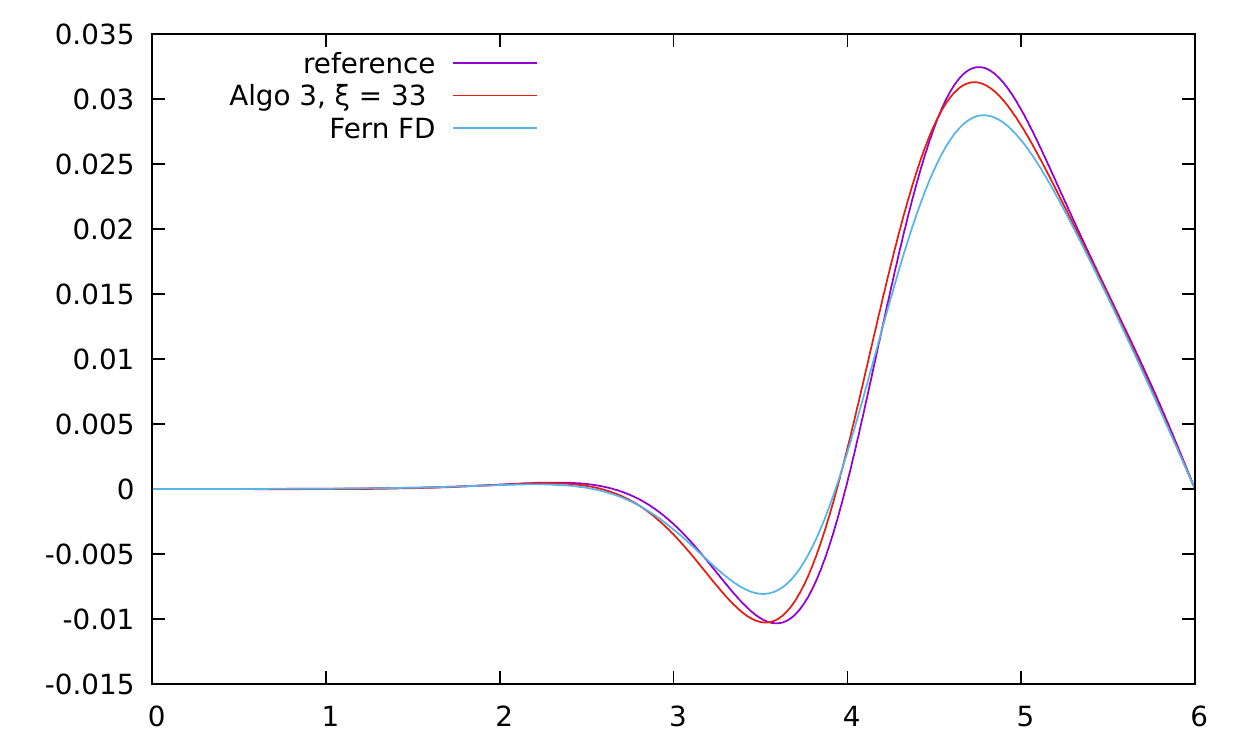}
    \caption{rate = 4}
    \label{fig:prj_prs-coret_Van-Kan_m3_r0_rate4_beta33}
  \end{subfigure}
\begin{subfigure}[h]{0.49\textwidth}
    \centering
    \includegraphics[width=\textwidth]{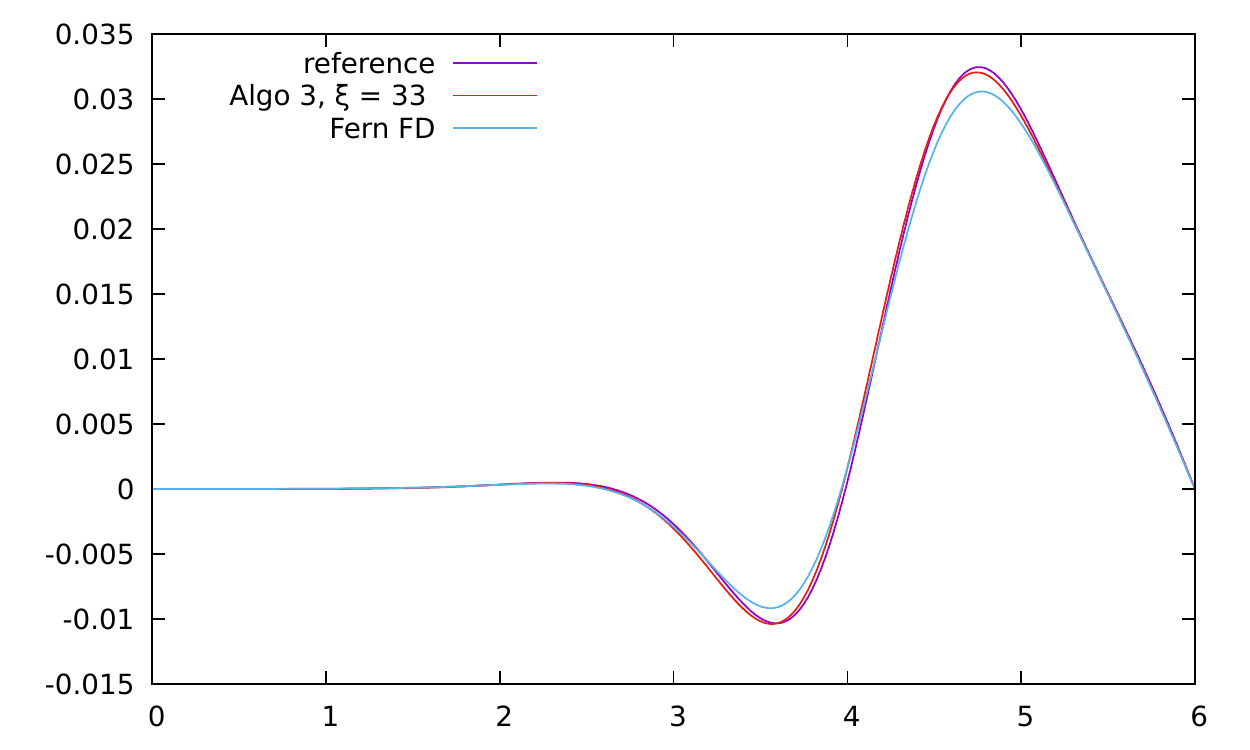}
    \caption{rate = 5}
    \label{fig:prj_prs-coret_Van-Kan_m3_r0_rate5_beta33}
  \end{subfigure}
  \caption{Structure displacement of Fernandez fully decoupled  scheme  (Fern FD) and Algorithm 3 (Algo 3) with $ \xi = 33 $ at final time}
  \label{fig:prj_prs-coret_Van-Kan_m3_r0_beta33}
\end{figure}